\documentclass[5p]{elsarticle}

\usepackage{graphicx}
\usepackage{float}
\DeclareGraphicsExtensions{.pdf}
\usepackage{amsmath,amsfonts,amsthm} % Math packages for equations
\usepackage{aas_macros}
\usepackage[T1]{fontenc}
\usepackage[utf8]{inputenc} % Required for inputting international characters
\usepackage{bm}
\usepackage[english]{babel} % English language hyphenation
\usepackage{acronym}
\usepackage{xspace}
\usepackage{footmisc}
\usepackage[hyphens]{url}
\usepackage[hidelinks,colorlinks=true]{hyperref}
\usepackage{siunitx}
\usepackage{lineno}
\usepackage{extdash}
\hypersetup{breaklinks=true}

\begin{document}

\title{A Hybrid Approach to Event Reconstruction for Atmospheric Cherenkov Telescopes Combining Machine Learning and Likelihood Fitting}

\author[1]{Georg Schwefer\corref{cor1}}
\ead{georg.schwefer@mpi-hd.mpg.de}
\cortext[cor1]{Corresponding author}
\author[2]{Robert Parsons}
\author[1]{Jim Hinton}

\affiliation[1]{organization={Max-Planck-Institut für Kernphysik},
addressline={Saupfercheckweg 1}, postcode={69117}, city={Heidelberg}, country={Germany}}
\affiliation[2]{organization={Institut f\"ur Physik, Humboldt-Universit\"at zu Berlin},
addressline={Newtonstr. 15}, postcode={12489}, city={Berlin}, country={Germany}}

\begin{abstract}
The imaging atmospheric Cherenkov technique provides potentially the highest angular resolution achievable in astronomy at energies above the X-ray waveband. High-resolution measurements provide the key to progress on many of the major questions in high energy astrophysics, including the sites and mechanisms of particle acceleration to PeV energies. The huge potential of the next-generation CTA observatory in this regard can be realised with the help of improved algorithms for the reconstruction of the air-shower direction and energy.

Hybrid methods combining maximum-likelihood-fitting techniques with neural networks represent a particularly promising approach and have recently been successfully applied for the reconstruction of astrophysical neutrinos.
Here, we present the \emph{Free\-PACT} algorithm, a hybrid reconstruction method for IACTs. In this, making use of the neural ratio estimation technique from the field of likelihood-free inference, the analytical likelihood used in traditional image likelihood fitting is replaced by a neural network that approximates the charge probability density function for each pixel in the camera.

The performance of this improved algorithm is demonstrated using simulations of the planned CTA southern array. For this setup, \emph{Free\-PACT} provides significant performance improvements over analytical
likelihood techniques, with improvements in angular and energy resolution of $25\%$ or more over a wide energy range and an angular resolution as low as $\ang{;;40}$ at energies above $50\,\mathrm{TeV}$ for observations at $20^{\circ}$ zenith angle.
It also yields more accurate estimations of the uncertainties on the reconstructed parameters and significantly speeds up the reconstruction compared to analytical likelihood techniques while showing the same stability with respect to changes in the observation conditions. Therefore, the \emph{Free\-PACT} method is a promising upgrade over the current state-of-the-art likelihood event reconstruction techniques.

\end{abstract}

\maketitle

% ----------------------------------------------------------------------------------------
% ----------------------------------------------------------------------------------------
% ----------------------------------------------------------------------------------------
\section{Introduction\label{sec:introduction}}

Imaging Atmospheric Cherenkov Telescope (IACT) arrays measure the flash of Cherenkov light emitted by the extensive air showers produced in interactions of high energy gamma rays in the atmosphere. In contrast to astronomy at most other wavelengths, the energy and direction of the individual gamma rays are reconstructed offline from the recorded Cherenkov images (or data-cubes, with time-resolved information). Thus, the performance of the reconstruction software is crucial and can be a limiting factor for IACTs. As the field of ground-based gamma-ray astronomy has developed, many different and ever-improving algorithms have been tested for both direction and energy reconstruction. The advent of the Cherenkov Telescope Array (CTA; \cite{CTAConsortium:2017dvg}) in particular has prompted a new wave of algorithm development (see e.g.~\citep{Miener:2021ixs,Jacquemont:2020buk}), to push toward the limits of achievable resolution (see e.g.~\cite{Hofmann:2006wf}). The current state-of-the-art techniques involve the use of image likelihood fitting. In these, one aims to estimate the shower parameters (direction, impact point, energy and depth of shower maximum) by maximising the likelihood of observing the measured images with respect to these parameters. The per-pixel charge likelihood is usually described as an analytical function based on the convolution of a poisson probability density function (PDF) with a gaussian photosensor resolution PDF that encompasses the `excess noise factor' of the photosensors and a baseline variance resulting from night sky background (NSB) fluctuations and electronic noise~\cite{deNaurois:2009ud}. The expectation value of this PDF at each pixel position is obtained from average-image templates, generated by making use of semi-empirical models of air-shower development~\citep{deNaurois:2009ud} or directly on Monte Carlo simulations of showers (\emph{ImPACT}; \cite{Parsons:2014voa}). These image templates are generated on a fixed grid of simulated parameters and a multidimensional interpolation is performed to find the expectation for any set of shower parameters.
The use of this template likelihood-fitting technique has significantly improved the performance of IACTs over the lifetime of the current generation of instruments, resulting in almost a factor of two improvement in angular resolution~\citep{Parsons:2014voa} over classical geometrical Hillas methods~\cite{AHARONIAN1997343}. However, ultimately one of the limitations of these techniques is the incomplete description of the per-pixel charge PDF by the analytical likelihood function used in the fit procedure.  

In practise, intrinsic shower deviations from the mean shower template for a given impact distance, shower energy and depth of shower maximum are highly non-Poissonian in nature, with fluctuations in the development of the shower contributing significantly. Similarly, the photosensor charge resolution function is not necessarily gaussian, but depends on the photosensor, electronics chain, and charge extraction algorithm. Whilst it is possible in principle to follow the template/look-up table approach also for the PDF (see for example~\cite{Joshi:2018qbo} for the related case of gamma-ray astronomy via ground-level particle detection), machine learning provides a very attractive alternative to a binned approach to the problem. This is especially true given the high dimensionality of the parameter space for the present problem, which effectively prohibits filling look-up tables with sufficient statistics to obtain smooth distributions. Such a hybrid reconstruction method combining machine learning and likelihood fitting was for example recently used to enhance the sensitivity of the IceCube Neutrino Observatory in its Cascade event channel, leading to the first observation of neutrinos from the Galactic plane~\citep{IceCube:2023ame}.

Here we adopt another hybrid machine learning\Hyphdash likelihood method that was also recently applied in the context of astrophysical neutrino telescopes~\cite{ELLER2023168011} and makes use of the neural ratio estimation technique from the field of likelihood-free inference~\cite{pmlr-v119-hermans20a}. We use the example of CTA to show how the gamma-ray reconstruction performance can be boosted with this \emph{Free\-PACT} (likelihood-free \emph{ImPACT}) approach.

In Section~\ref{sec:approach} we describe the \emph{ImPACT} and \emph{Free\-PACT} image likelihood-fitting reconstruction approaches. The data sets used for training and testing and the performance of the reconstruction methods are summarised in Section~\ref{sec:perf}. In Section~\ref{sec:discuss} we discuss broader implications and practical aspects of adoption of the method for CTA and other observatories. We conclude in Section~\ref{sec:summary}.

\section{Image Likelihood-fitting Methods}
\label{sec:approach}

\subsection{General Formulation} % 
\label{sec:general_formulation}
In general, in image likelihood-fitting reconstructions for IACTs, one typically aims to estimate (a subset of) the six shower parameters energy $E$, source direction $\vec{r}_{\rm src}$, shower core location $\vec{r}_{\rm core}$ and $X_\mathrm{max}$ (the atmospheric slant depth at which the maximum number of Cherenkov-light emitting particles is reached). For a given, fixed pointing direction and under the assumption of independent per-pixel PDFs, the likelihood of observing images of charge $c_{ij}$ in pixels $j$ of telescopes $i$ can generally be written as 
\begin{equation}
\label{eq:total_likelihood}
    \mathcal{L}=\prod_{i=1}^{N_\mathrm{tel}}\prod_{j=1}^{N_\mathrm{pix}}p(c_{ij}|\vec{r}_{\rm pix}^{\;ij},\vec{r}_{\rm tel}^{\;i},\vec{\eta}).
\end{equation}
Here, $\vec{r}_{\rm tel}$ is the telescope position on the ground, $\vec{r}_{\rm pix}$ is the pixel coordinate and $\vec{\eta}$ denotes the six shower parameters to be estimated. $\vec{r}_{\rm src}$ and $\vec{r}_{\rm pix}$ are typically both chosen in a common sky coordinate frame that is centered on the array pointing direction (called the \emph{nominal frame} or field-of-view coordinate system).

Typically, while used for the evaluation and eventual maximisation of the total likelihood $\mathcal{L}$, this set of parameters is not used directly for the construction and formulation of the per-pixel likelihood $p$. Instead, making use of the (approximate) symmetries of the camera, a transformed set is adopted: First, the pixel coordinates are translated into new coordinates $\vec{r}_{\mathrm{trans}}=\vec{r}_{\rm pix}-\vec{r}_{\rm src}$ such that the source position lies in the centre of the camera. This neglects the degradation of the camera optics towards the edge of the field of view, but, as we show in Figure~\ref{fig:offset_dependence}, the final performance of the algorithm is only weakly dependent on the field-of-view offset even in the worst-case scenario. Then, making use of the approximate rotational symmetry of the telescopes and cameras, these coordinates are rotated into another set of coordinates $\vec{\Tilde{r}}_{\rm pix}$ such that the shower axis lies along the $x$-axis of the nominal frame. In the final step, the core location and telescope position are projected to the plane perpendicular to the shower axis to obtain the (projected) shower impact distance $d$. The per-pixel likelihood is then formulated as a function of five quantities in total: $E$, $X_{\rm max}$, $d$, and $\vec{\Tilde{r}}_{\rm pix}$.

Different implementations of image likelihood-fitting reconstructions will then typically differ only in the choice of the per-pixel likelihood $p(c|E, X_{\rm max}, d, \vec{\Tilde{r}}_{\rm pix})$. In the following sections, we will describe the respective implementations for the \emph{ImPACT} and \emph{Free\-PACT} methods.

\subsection{ImPACT}
As mentioned in Section~\ref{sec:introduction}, in the current state-of-the-art methods such as \emph{ImPACT}, the per-pixel PDF $p$ is commonly described analytically following~\cite{deNaurois:2009ud} as
\begin{equation}
\label{eq:full_convolved_likelihood}
\begin{aligned}
    &p(c|\mu(E, X_{\rm max}, d, \vec{\Tilde{r}}_{\rm pix}),\sigma_{\rm p},\sigma_{\rm \gamma})=\\
    &\sum_{n} \frac{\mu^ne^{-\mu}}{n!\sqrt{2\pi(\sigma_{\rm p}^2+n\sigma_{\rm \gamma}^2)}}\;\exp{\left(-\frac{(c-n)^2}{2(\sigma_{\rm p}^2+n\sigma_{\rm \gamma}^2)}\right)},\\
\end{aligned}
\end{equation}
which for $\mu\gg0$ simplifies to 
\begin{equation}
\label{eq:gaussian_likelihood}
\begin{aligned}
    &p(c|\mu(E, X_{\rm max}, d, \vec{\Tilde{r}}_{\rm pix}),\sigma_{\rm p},\sigma_{\rm \gamma})=\\
    &\frac{1}{\sqrt{2\pi(\sigma_{\rm p}^2+\mu(1+\sigma_{\rm \gamma}^2))}}\;\exp{\left(-\frac{(c-\mu)^2}{2(\sigma_{\rm p}^2+\mu(1+\sigma_{\rm \gamma}^2))}\right)}.\\
\end{aligned}
\end{equation}
In these, $\sigma_{\rm \gamma}$ and $\sigma_{\rm p}$ are measures for the RMS of the single photo-electron (P.E.) distribution and the RMS of the charge PDF in the absence of any Cherenkov light, known as the pedestal, respectively. These can be extracted either from camera data or simulations. For \emph{ImPACT}, the value of $\mu(E, X_{\rm max}, d, \vec{\Tilde{r}}_{\rm pix})$ is taken from interpolating a grid of average-image templates generated from Monte Carlo simulations. Henceforth, the distributions in equations \ref{eq:full_convolved_likelihood} and \ref{eq:gaussian_likelihood} will be referred to as the convolved and Gaussian \emph{ImPACT} PDFs, respectively.

\subsection{FreePACT}

For \emph{Free\-PACT}, we use the neural ratio estimation method from the field of likelihood-free inference~\cite{pmlr-v119-hermans20a}. Our implementation is similar to that described in~\cite{ELLER2023168011} for the reconstruction of astrophysical neutrinos. We briefly outline the core points of the method and the application to our problem below and refer to the original publications for more details.

In general, the goal of neural ratio estimation is to compute the likelihood-to-evidence ratio
\begin{equation*}
    r(\vec{x}|\vec{\theta})=\frac{p(\vec{x}|\vec{\theta})}{p(\vec{x})}
\end{equation*}  for a quantity (or a set of quantities) $\vec{x}$ given a set of parameters $\vec{\theta}$. As the evidence is independent of $\vec{\theta}$, minimising $-\log r(\vec{x}|\vec{\theta})$ is equivalent to minimising $-\log p(\vec{x}|\vec{\theta})$. It can be shown~\citep{Cranmer:2015bka,pmlr-v119-hermans20a} that $r(\vec{x}|\vec{\theta})$ can be calculated from the output of a particular binary classifier $0<d(\vec{x},\vec{\theta})<1$ as 
\begin{equation*}
    r(\vec{x}|\vec{\theta})=\frac{d(\vec{x},\vec{\theta})}{1-d(\vec{x},\vec{\theta})}.
\end{equation*}
This classifier must be trained with binary cross-entropy to differentiate between samples drawn from the joint PDF $p(\vec{x},\vec{\theta})$ (class 1) and samples drawn from the product of the marginal PDFs $p(\vec{x})p(\vec{\theta})$ (class 0).

Translating this to our application following equation~\ref{eq:total_likelihood}, $\vec{x}$ represents the measured charge $c$ and $\vec{\theta}$ represents the five parameters $E$, $X_{\rm max}$, $d$, and $\vec{\Tilde{r}}_{\rm pix}$. The samples from the joint distribution $p(c, E, X_{\rm max}, d, \vec{\Tilde{r}}_{\rm pix})$ are just simulated Monte Carlo events, the samples from the product of marginals $p(c)p(E, X_{\rm max}, d, \vec{\Tilde{r}}_{\rm pix})$ can be obtained from Monte Carlo events by randomly shuffling the charge values between events.

\begin{figure}
\centering
\includegraphics[trim={4cm 3cm 6cm 2.2cm},clip,width=0.5\textwidth]{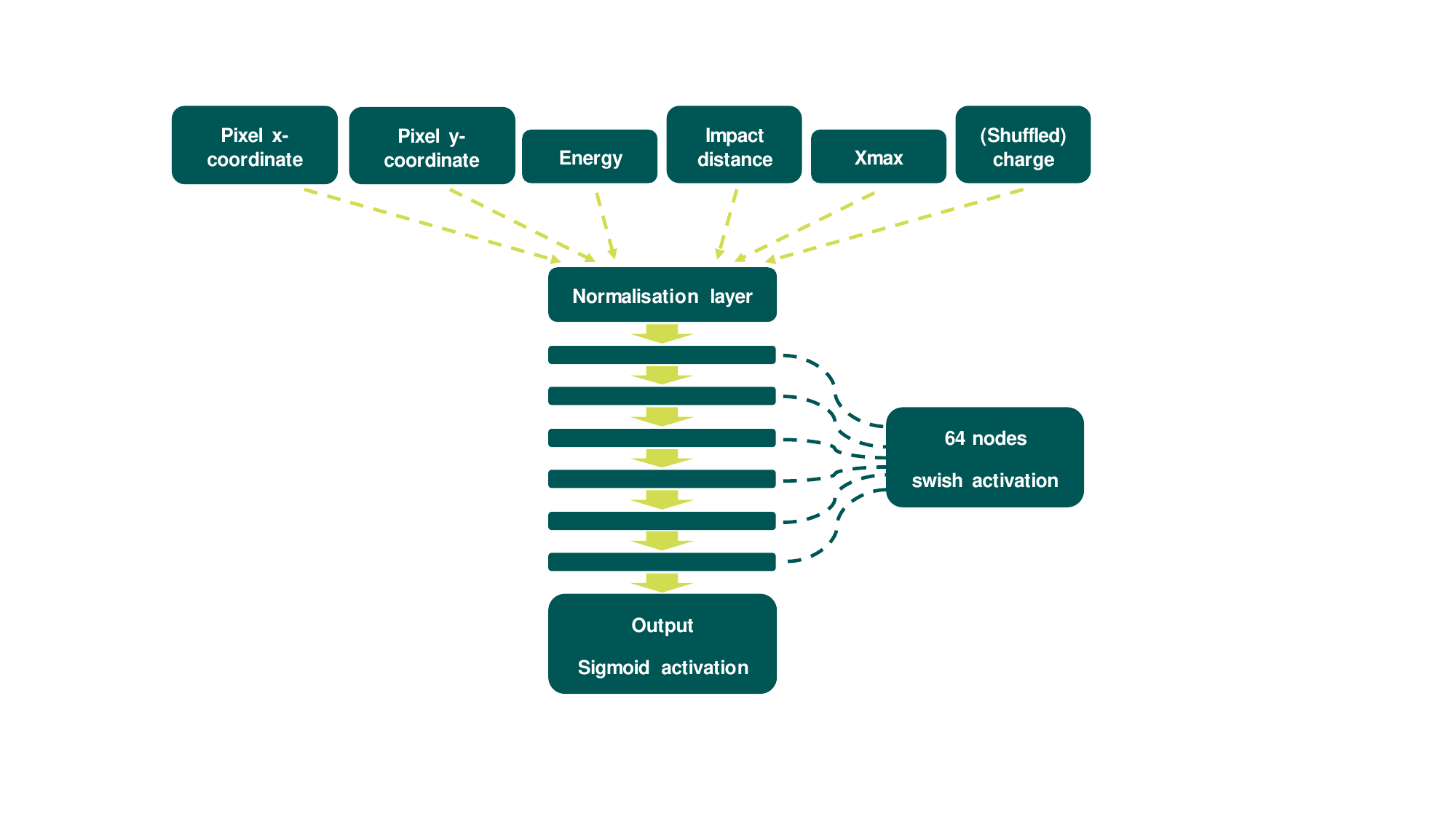}
\caption{
Schematic depiction of the neural network architecture used for the \emph{Free\-PACT} models
}
\label{fig:nn_design}
\end{figure}

We implement the classifier at the core of neural ratio estimation and therefore the \emph{Free\-PACT} algorithm as a dense feed-forward neural network. The architecture follows that proposed in~\cite{ELLER2023168011} and features six input nodes, a layer for normalising and scaling the input data, six hidden layers and a single output node (see Figure \ref{fig:nn_design}). Most notably in the normalising layer, as the measured charge varies over multiple orders of magnitude, but can also attain negative values, values are scaled linearly between $-20\,\mathrm{P.E.}$ and $20\,\mathrm{P.E.}$ and logarithmically outside, keeping the sign. The value of $20\,\mathrm{P.E.}$ is chosen to fully encompass the noise-dominated regime in the linear scaling and the high-charge tail of the distribution in the logarithmic scaling. In the same step, we also take the logarithm of the gamma-ray energy to account for its variation over four orders of magnitude. For the hidden layers, we have tested a number of architectures differing in depth and number of nodes per layer and have found the training results to be robust against these variations. To minimise training and evaluation time, we have finally opted for a comparatively small network with 64 nodes in each of the six hidden layers. With this architecture, the model has around $2\cdot10^4$ trainable parameters, a very modest number, especially compared to reconstruction methods based on convolutional neural networks (CNNs) that can have more than $10^6$ trainable parameters (see e.g.~\cite{xie2019utterance} for the model used in~\cite{Miener:2021ixs}). As in~\cite{ELLER2023168011}, we use the \textit{swish} activation function for the hidden layers to obtain a smooth neural network output and thus a differentiable likelihood-to-evidence ratio. During training, we apply a sigmoid activation to the output layer. This is changed to a linear activation for the evaluation of the model, with which the model directly outputs the logarithm of the likelihood-to-evidence ratio~\citep{ELLER2023168011}. A separate model is trained for every telescope type, i.e. every combination of camera type and telescope optical properties.

\section{Performance}
\label{sec:perf}
In this section we will describe the training of the \emph{Free\-PACT} models, the generation of the \emph{ImPACT} templates used for comparison and evaluate their performance on Monte Carlo simulations.

All simulations -- the \emph{Free\-PACT} model training data, the input to the \emph{ImPACT} templates and the set of test simulations to evaluate the performance of the reconstruction -- are generated using the CORSIKA/sim\textunderscore telarray Monte Carlo simulation suite~\citep{Heck:1998vt,Bernlohr:2008kv}. For this study, we focus on the {\it Alpha} configuration of the CTA southern array to be built in Paranal, Chile~\cite{alpha_config}. This array configuration consists of 14 Medium-Sized Telescopes (MSTs) equipped with the FlashCam camera~\citep{Puhlhofer:2019gom} and 37 Small-Sized Telescopes (SSTs) equipped with a SiPM-based camera based on the CHEC-S prototype design~\citep{Depaoli:2023fwp}. A detailed overview of the layout can also be found in~\ref{app:array_layout}. The site and telescope models follow a preliminary version of the CTA {\it Prod6} simulation configuration. For the previous officially released simulation configuration of CTA, {\it Prod5}, please see~\cite{bernlohr_2022_6218687}.

The processing of the training data and the reconstruction of the test simulations is done using a modified v0.20 of the open-source \emph{ctapipe} framework that has been written for CTA~\citep{Linhoff:2023atg,karl_kosack_2023_8335474}.  For the likelihood fit, we use the \emph{iminuit} package~\citep{hans_dembinski_2022_7115916} based on the \emph{MINUIT} minimiser~\cite{James:1975dr}. Angular and energy resolution curves are calculated using v0.10.1 of the open-source \emph{pyirf} package ~\citep{maximilian_linhoff_2023_8348922} and v1.1 of the open-source \emph{gammapy} package~\cite{gammapy:2023, gammapy:zenodo-1.1}.

\subsection{FreePACT Model Training}
\label{sec:nn_training}

For each simulated gamma-ray shower, the charge images of every triggered telescope are extracted from the simulated waveform data-cubes using the \textit{FlashCamExtractor} (MST) and \textit{NeighborPeakWindowSum} (SST) algorithms as implemented in \emph{ctapipe}, respectively. The resulting images are then cleaned using a tail-cuts cleaning method with thresholds of $10\,\mathrm{P.E.}$ and $5\,\mathrm{P.E.}$. As we have found this to improve the quality of the reconstruction, we also use the two rows of pixels around the cleaned image both for the training data and during reconstruction.
We use simulations of an on-axis point source gamma-ray source for the training. As we show in Figure~\ref{fig:offset_dependence}, the resulting models also perform reasonably well at large field-of-view offsets. Furthermore, we also found no performance improvements when using diffuse gamma-ray simulations instead.
The total training data consists of $\mathcal{O}(10^7)$ pixels. We have found a good performance of the model across the entire energy range when an approximately equal number of these pixels come from simulations with primary gamma-ray spectra $\propto E^{-1}$ and $\propto E^{-2}$. However, this should not be seen as a strict requirement.  Creating this data set requires only a modest number of simulated gamma-ray air showers. We have verified that the performance of the model is not strongly dependent on the size of the training set.

The models are trained until convergence is reached which is typically the case after about 50 epochs. We use the Adam optimiser~\citep{kingma2017adam} and a learning rate of 0.001. On 10 CPU cores, the training only takes a few minutes.

\subsection{ImPACT template generation}
\label{sec:impact_templates}
\emph{ImPACT} templates are generated using 2 different me\-thods: For the first, classical, method, sets of dedicated simulations are created for each telescope type. Each of these sets features simulations of gamma rays at a fixed energy and core location. The simulated array is a line of 30 equidistant telescopes extending for $1.2\,\mathrm{km}$ from the core location. This configuration is chosen to effectively 
populate the grid in shower parameter space on which the templates are created. After image extraction and cleaning with the methods described in Section \ref{sec:nn_training}, the template creation from these simulations is done using the \emph{template\textunderscore builder} code\footnote{\url{https://github.com/ParsonsRD/template_builder}}. Operationally, the fact that this method requires such special simulations is a significant disadvantage compared to the \emph{Free\-PACT} method that uses standard array simulations.

 The second method to generate \emph{ImPACT} templates makes use of the \emph{Free\-PACT} models. As demonstrated in figures \ref{fig:mst_charge_distribution} and \ref{fig:sst_charge_distribution}, it is possible to calculate the charge PDF on a grid of pixel coordinates and shower parameters from the \emph{Free\-PACT} models. Thus, \emph{ImPACT} templates can be generated from these models using the expectation value of the PDFs. This shows that the generation of \emph{ImPACT} templates is actually just a corollary of the \emph{Free\-PACT} approach. These templates are however mostly of interest for the purpose of comparison to the full \emph{Free\-PACT} algorithm, as they do not exploit the main feature of the \emph{Free\-PACT} method, i.e. the fact that an analytical likelihood function is no longer necessary. They do however allow us to isolate the source of performances differences between \emph{ImPACT} and \emph{Free\-PACT}. 

Finally, we determine the values of the free parameters $\sigma_{\rm \gamma}$ and $\sigma_{\rm p}$ in the likelihood functions in equations \ref{eq:full_convolved_likelihood} and \ref{eq:gaussian_likelihood}. $\sigma_{\rm \gamma}$ is calculated from the same single P.E. distribution for each camera that is also an input to the telescope simulations. $\sigma_{\rm p}$ is calculated from simulated air shower events by taking the standard deviation of the distribution of measured charge in pixels without Cherenkov light for default dark sky, i.e. moonless, observation conditions. We find values of $\sigma_{\rm \gamma}=0.5\,\mathrm{P.E.}$ and $\sigma_{\rm p}=1.6\,\mathrm{P.E.}$ for the MST and $\sigma_{\rm \gamma}=0.2\,\mathrm{P.E.}$ and $\sigma_{\rm p}=0.8\,\mathrm{P.E.}$ for the SST. As a test, we have also varied these parameters over a wide range of values have found no difference in performance.

\subsection{Charge PDF comparisons}
\label{sec:charge_pdf_comparison}

On a lower level than the instrument response functions (IRFs) and corresponding resolution curves, it is possible to evaluate the performance of the \emph{Free\-PACT} method by comparing the per-pixel charge PDFs $p(c|E, X_{\rm max}, d, \vec{\Tilde{r}}_{\rm pix})$ (c.f. Section~\ref{sec:general_formulation}) for a given set of shower parameters as predicted by \emph{Free\-PACT} and \emph{ImPACT} to the true distributions.

These true distributions are obtained from simulating a set of events at a given fixed energy and impact distance from a single telescope and then taking the per-pixel histogram of measured charges over the set of events. Just like the simulations used to train the \emph{Free\-PACT} models, these simulations are also generated with on-axis gamma rays. Any comparison of \emph{Free\-PACT} models trained on off-axis simulations to the true distribution obtained from the corresponding off-axis simulations should yield very similar results. Furthermore, as shown in Figure~\ref{fig:offset_dependence}, the models trained on on-axis simulations also perform reasonably well at large field-of-view offsets. This is a clear indication that the true distributions will not look markedly different for off-axis gamma rays.

The \emph{Free\-PACT} distributions are obtained by multiplying the likelihood-to-evidence ratio obtained from the neural network for the simulated shower parameters and pixel positions with the total charge distribution from the training data set (i.e. the evidence). Due to the large dimensionality of the parameter space, this is done only at a few points as a cross-check.

\begin{figure*}[!ht]
\centering
\includegraphics[width=0.49\textwidth]{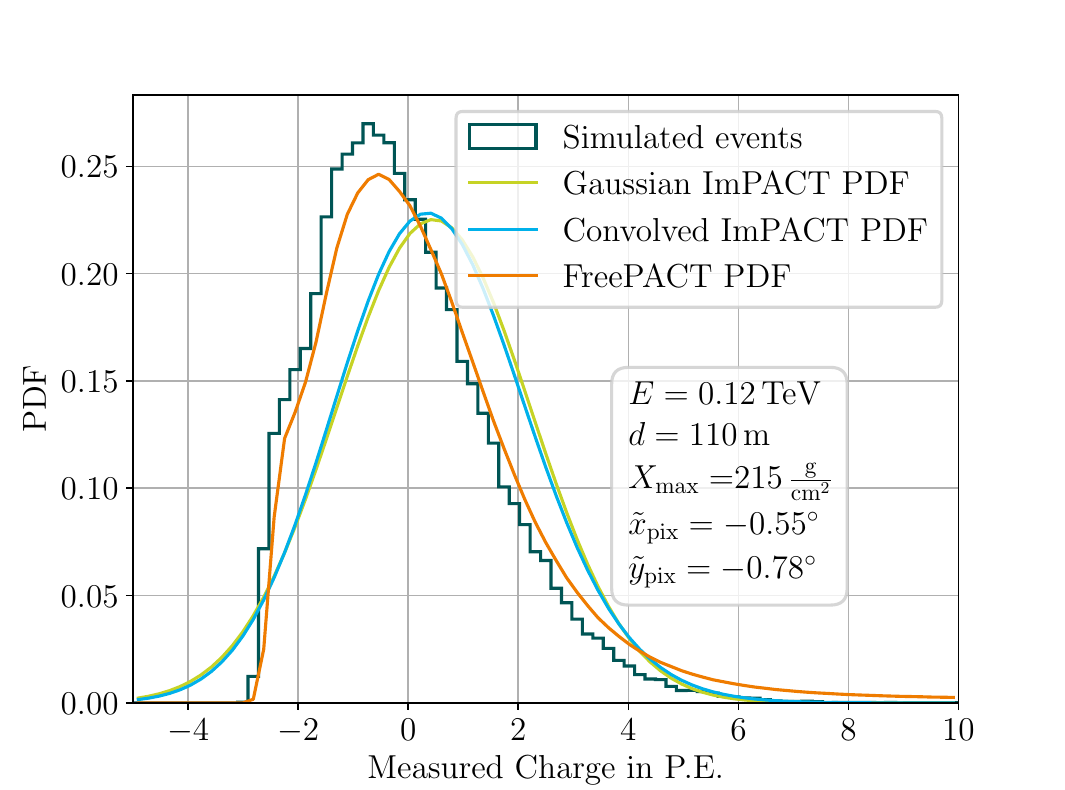}
\includegraphics[width=0.49\textwidth]{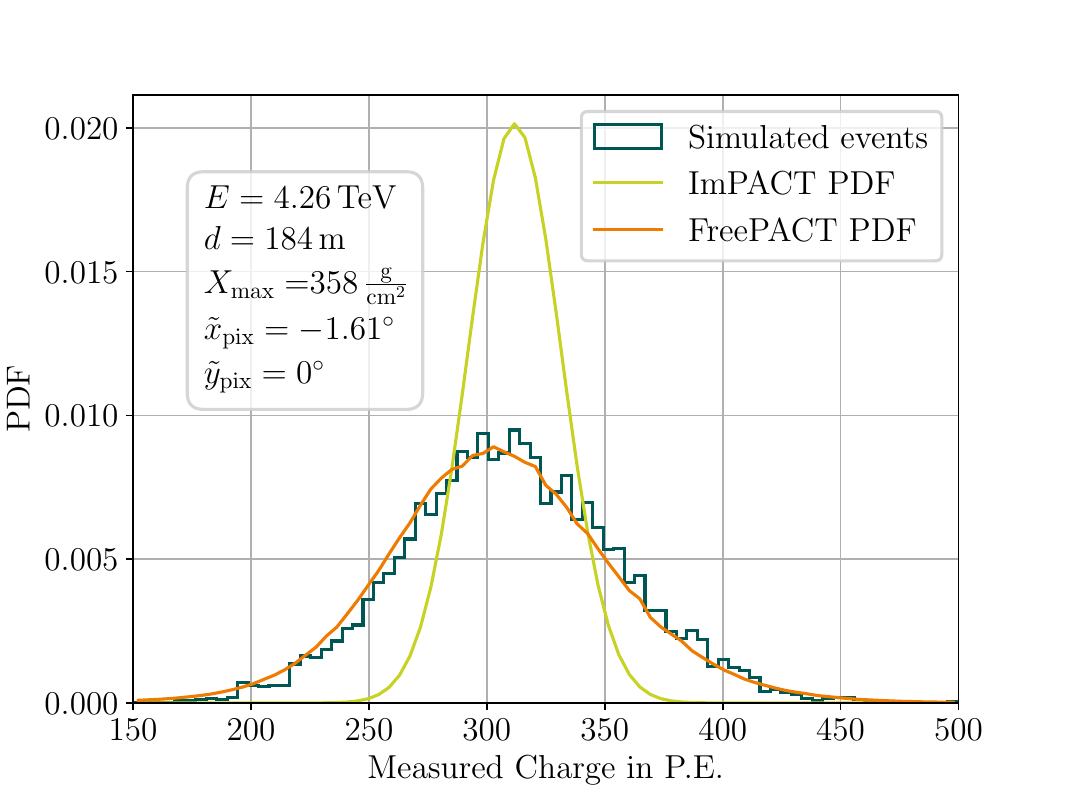}
\caption{
Charge distribution at the given pixel position for the given shower parameters in a FlashCam-MST with the analytical \emph{ImPACT} charge PDFs and the \emph{Free\-PACT} distribution. On the left, the charge is dominated by noise photons, on the right, the charge is dominated by Cherenkov photons. On the right, as the Gaussian and convolved \emph{ImPACT} PDFs are identical, we only show one curve.
}
\label{fig:mst_charge_distribution}
\end{figure*}

In Figure~\ref{fig:mst_charge_distribution}, we show two example distributions for the MST. Corresponding comparisons of SST PDFs can be found in Figure~\ref{fig:sst_charge_distribution} in~\ref{app:sst_charge_pdf}. The specific shower parameters and pixel positions used are given in each panel of Figure~\ref{fig:mst_charge_distribution}. All showers with a value of $X_{\rm max}$ within $12.5\,\mathrm{g}\:\!\mathrm{cm}^{-2}$ of the given value are considered.  

In the left panel of Figure~\ref{fig:mst_charge_distribution}, we show a charge distribution that is dominated by NSB photons, the right panel contains a distribution dominated by shower Cherenkov light. For the NSB-dominated pixel, we find that the \emph{Free\-PACT} PDF is able to capture the specific shape of the noise distribution resulting from the photosensor and electronics chain properties of the camera as well as the charge extraction algorithm. Naturally, the Gaussian \emph{ImPACT} PDF can not fully describe such a shape, but neither can the convolved \emph{ImPACT} PDF that is specifically formulated to model the charge distribution at low intensities. 

At larger image intensities the \emph{ImPACT} PDF is shown to be significantly narrower than that from \emph{Free\-PACT}. The high intensity events represent the regime where shower-to-shower fluctuations become the dominant effect in the PDF. As these effects are not included in the \emph{ImPACT} PDF, it rather poorly reflects the true distribution of pixel amplitudes.
Similar conclusions can be drawn from additional comparisons elsewhere in the parameter space as well as from analogous tests with the SST (see Figure~\ref{fig:sst_charge_distribution}). As the addition of an improved PDF description represents the only change in \emph{Free\-PACT} from the standard \emph{ImPACT} reconstruction it can therefore be identified as the origin of the reconstruction performance improvements presented in Section~\ref{sec:ang_E_res}.

 As already stated above, these results are robust against changes in the values of the parameters $\sigma_{\rm p}$ and $\sigma_{\rm \gamma}$. Of course, it is in principle always possible to further modify the functional form of the \emph{ImPACT} PDF as given in equation~\ref{eq:full_convolved_likelihood}: For example, one could introduce another uncertainty parameter $\sigma_{\rm sts}$ that scales proportional to $n$ to try to account for the shower-to-shower fluctuations (see~\cite{Holler:2015tca} for an example of such a modification). However, it is unclear whether any analytical functional form of this nature actually manages to grasp the full complexity of the distribution over the entire parameter space, as it will always be more limited than the \emph{Free\-PACT} distribution. A growing number of parameters also defeats the main advantage of the analytical form, namely its intuitive understanding.

\subsection{Event Selection and Analysis Configurations}
\label{sec:event_selection}
In order to assess the reconstruction performance of the \emph{Free\-PACT} algorithm, we use sets of test gamma-ray simulations as described above. For these, the camera images are extracted and cleaned with the same methods used for the \emph{Free\-PACT} model training data described in Section \ref{sec:nn_training} and summary parameters including the Hillas parameters~\citep{1985ICRC....3..445H} are calculated. For the reconstruction step, we consider four sets of analyses.

\begin{itemize}
    \item Hillas/Random Forest: The directional reconstruction is performed with a standard geometrical Hillas method~\citep{AHARONIAN1997343}. Then, for the energy reconstruction, we train a Random Forest regression algorithm as implemented in \emph{ctapipe} on 15 image summary and geometry reconstruction parameters. The results from this analysis also serve as seeds for the \emph{ImPACT} and \emph{Free\-PACT} likelihood fits.
    \item ImPACT: This method uses the \emph{ImPACT} templates generated with the classical method described in Section \ref{sec:impact_templates} together with the full convolved analytical pixel likelihood. 
    \item ImPACT from FreePACT: This method uses the \emph{ImPACT} templates generated from the mean of the \emph{Free\-PACT} models as described in Section \ref{sec:impact_templates} together with the full convolved analytical pixel likelihood.
    \item FreePACT
\end{itemize}

 We do not compare our results to an analysis using reconstruction methods based on CNNs or similar algorithms. This is for two reasons: First, while showing great promise for classification tasks such as gamma-hadron separation (e.g.~\cite{Glombitza:2023qen}), such methods for current-generation instruments like H.E.S.S. have not been able to outperform likelihood-based methods, i.e. \emph{ImPACT}, in reconstruction tasks~\cite{Shilon:2018xlp}. The second reason is that there is currently no fully developed framework for a stereoscopic CNN-based analysis for CTA that would actually enable a fair comparison to our results. Frameworks that are currently under development such as \emph{CTLearn}~\citep{Miener:2021ixs} are at present not able to significantly outperform classical methods.
Unless specified differently, to ensure good quality images are included in the fit, we select only events that have at least four images with:
\begin{itemize}
    \item More than three pixels surviving image cleaning
    \item A total reconstructed intensity greater than 50 P.E.
    \item Less than $10\%$ of the image intensity in the two outermost pixel layers in the camera.
    \item The Hillas centroid within $3^{\circ}$ of the camera center
\end{itemize}

The first two cuts are used to assure there are sufficiently bright images in the event, whereas the latter two cuts assure only images well contained within the camera are used for reconstruction.

For all reconstruction methods, only the images passing these cuts are used for the reconstruction even if there are more triggered telescopes in the event.
For the results shown in this section, we do not apply gamma-hadron separation. That is because we want to stay independent of any particular gamma-hadron separation algorithm for the evaluation of the pure reconstruction performance. For performance curves of the full CTA southern array after gamma-hadron separation with a Random Forest classifier see \ref{app:rel_curves_gh}.  

As a reference to compare the event selection and the resolution performance of the different analyses outlined above against, we use the official CTA {\it Prod5} IRFs~\citep{cherenkov_telescope_array_observatory_2021_5499840}. For these, the source direction was reconstructed using a Boosted Decision Tree (BDT) that estimates the displacement between image centroid and source position along the shower axis in field-of-view coordinates~\citep{Lessard:2000yf}. Energy reconstruction and gamma-hadron separation were similarly done using a BDT algorithm.
It must be noted that the comparison of these IRFs to those produced in this work comes with an important caveat: The official CTA {\it Prod5} IRFs are based on event selection criteria different from the ones used here. In particular, they include energy-dependent cuts on the minimum telescope multiplicity and the gamma-hadron separation score that are optimised for point source sensitivity. Therefore, we do not expect the resolution of any of the analyses outlined above -- including the Hillas/Random Forest analysis -- to reproduce the resolution of the official CTA {\it Prod5} IRFs. Still, they serve as a useful reference to benchmark and compare our results against.

\begin{figure}[!ht]
\centering
\includegraphics[width=0.5\textwidth]{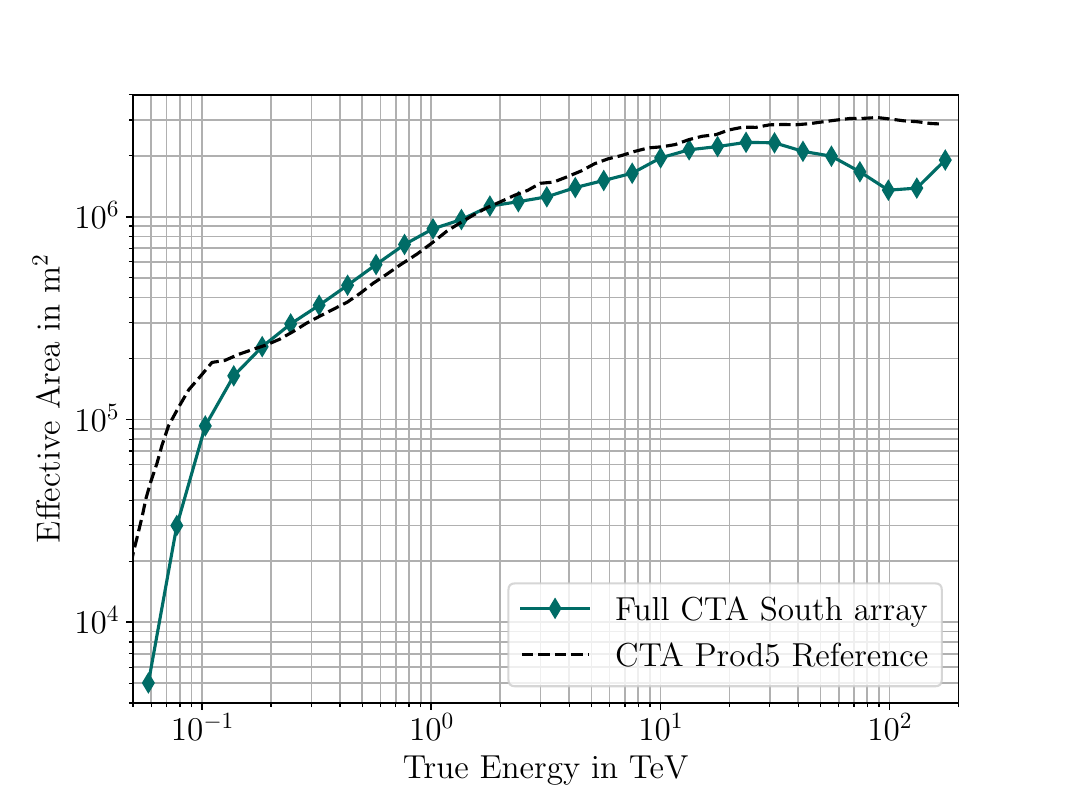}
\caption{
On-axis effective area of the event selection described in Section \ref{sec:event_selection} at $20^\circ$ zenith angle for the full CTA southern {\it Alpha} configuration array. This effective area is calculated before any gamma-hadron separation. For comparison, the black dashed curve shows the corresponding official CTA {\it Prod5} effective area which uses a different event selection and is calculated after gamma-hadron cuts are applied.
}
\label{fig:cta_south_aeff}
\end{figure}

The on-axis effective area of the event selection outlined above at $20^\circ$ zenith angle as well as the corresponding  official CTA {\it Prod5} reference curve are shown in Figure~\ref{fig:cta_south_aeff}. Overall, there is good qualitative agreement between the {\it Prod5} reference effective area and that for the analyses developed for this work. The remaining quantitative differences, in particular at the edges of the energy range, reflect the differences in event selection criteria.

\subsection{Angular and Energy Resolution}
\label{sec:ang_E_res}
\begin{figure}[!ht]
\centering
\includegraphics[trim={0cm 3cm 0cm 4.8cm},clip,scale=0.49]{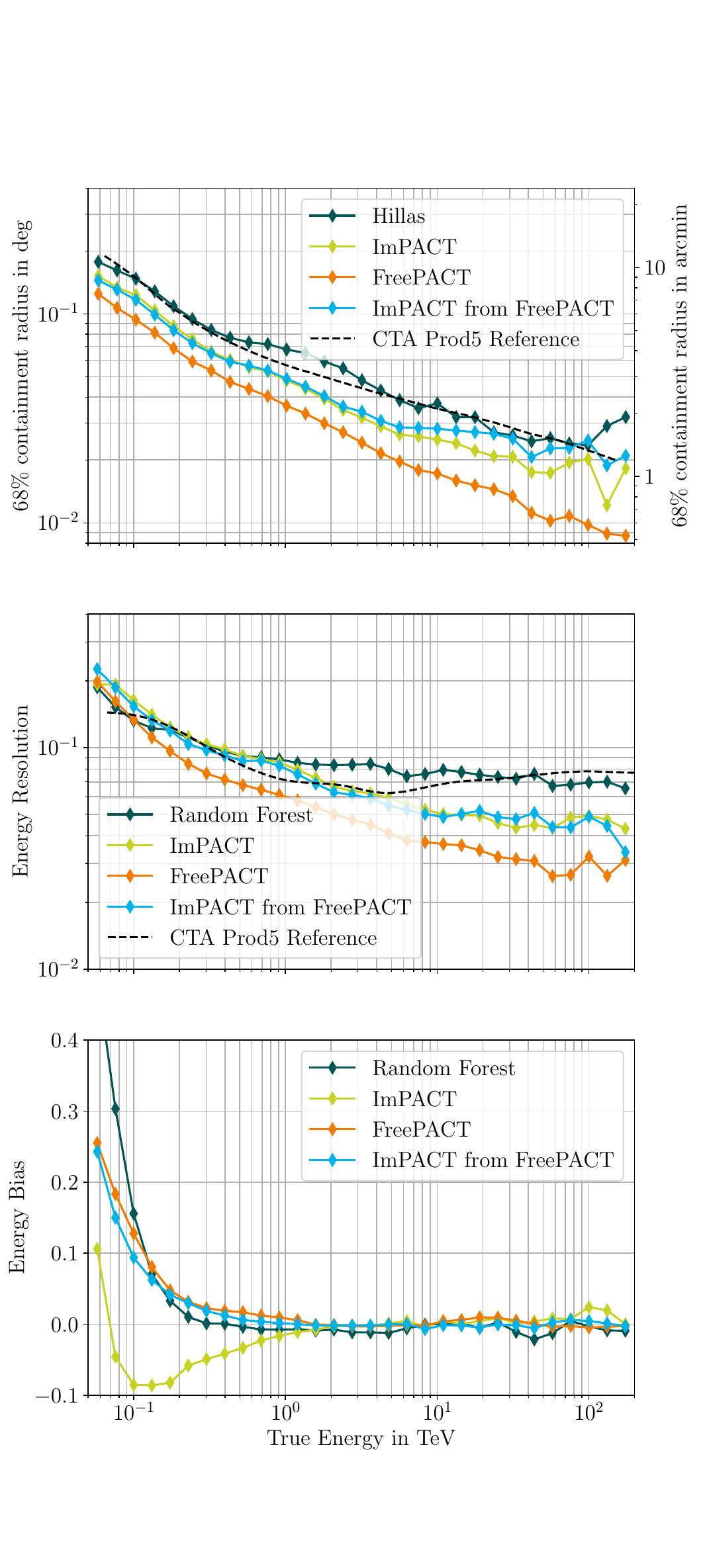}

\caption{
Angular resolution, energy resolution, and energy bias as a function of true gamma-ray energy for the CTA southern array for on-axis photons at $20^{\circ}$ zenith angle. Shown in dark green is the standard Hillas/Random Forest analysis that serves as a seed to the image likelihood fits. In lime green, we show the results from a conventional \emph{ImPACT} analysis. The \emph{Free\-PACT} resolution curves are shown in orange. The light blue curve shows the resolution achieved with an \emph{ImPACT} analysis with templates generated from the \emph{Free\-PACT} models as described in Section \ref{sec:impact_templates}. The official CTA Prod5 curve (after gamma-hadron separation) is shown in black for reference.
}
\label{fig:full_cta_resolution curves}
\end{figure}

In Figure~\ref{fig:full_cta_resolution curves}, we show the performance curves for on-axis gamma rays at $20^\circ$ zenith angle for the full CTA southern {\it Alpha} configuration array. The corresponding performance curves for the MST and SST sub-arrays are shown in figures \ref{fig:mst_resolution curves} and \ref{fig:sst_resolution curves} in \ref{app:subarray_performance}. Angular resolution is defined as the angular distance from the source position that contains $68\%$ of reconstructed events. Energy resolution and bias are defined as the half of the central $68\%$ interval and mean of the distribution of $\frac{E_{\rm reco}-E_{\rm true}}{E_{\rm true}}$, respectively. 
We also once again compare the resolution curves to the official CTA {\it Prod5} IRFs ~\citep{cherenkov_telescope_array_observatory_2021_5499840}.

We find the \emph{Free\-PACT} method to perform significantly better than all other algorithms for both angular and energy resolution and both telescope types over essentially the entire energy range. At $300\,\mathrm{GeV}$, the angular and energy resolution are improved\footnote{The relative improvement of \emph{Free\-PACT} over a reference resolution are calculated as $(\mathrm{reference}/\emph{Free\-PACT}) - 1$ } by more than $25\%$ with respect to both the Hillas/Random Forest and \emph{ImPACT} methods. At $30\,\mathrm{TeV}$, energy and angular resolution is improved by $\approx100\%$, i.e. a factor 2, compared to the Hillas/Random Forest method. Relative to \emph{ImPACT} we find improvements of $60\%$ for angular resolution and $40\%$ for energy resolution at the same energy.

The energy reconstruction bias is below $5\%$ for energies above $200\,\mathrm{GeV}$ and below $1\%$ for energies above $1\,\mathrm{TeV}$ for all methods. Towards the low-energy threshold, we observe the typical upwards bias as only events that are unusually bright for their energy manage to trigger the array.

Regarding the resolution performance of the \emph{ImPACT} methods, it is interesting to note that there are only small differences between the classically generated templates and those generated from the \emph{Free\-PACT} models. The difference in energy bias between the two methods at low energies can be traced back to a slight difference in the normalisation of the templates around the positions of the brightest pixels in a given image. This shows that the performance gained using the \emph{Free\-PACT} method really comes from correctly modelling the \emph{shape} of the charge PDFs around the mean value. It also shows that the process of generating the \emph{ImPACT} templates can indeed become a simple byproduct of the \emph{Free\-PACT} method.

The comparison to the official CTA {\it Prod5} reference resolution curves further illustrates the improvement in precision achieved with the \emph{Free\-PACT} method as the reference resolution curves are qualitatively similar to those of the Hillas/Random Forest analysis. Due to the difference in event selection criteria between the analyses developed for this work and the {\it Prod5} reference, we refrain from any further quantitative comparison here.

These conclusions also qualitatively hold for observations at different zenith angles. As an example of this, we show in Figure~\ref{fig:full_cta_resolution curves_z50} in \ref{app:fifty_zenith} the performance curves for on-axis observations at a zenith angle of $50^\circ$. 

\begin{figure}
\centering
\includegraphics[scale=0.49]{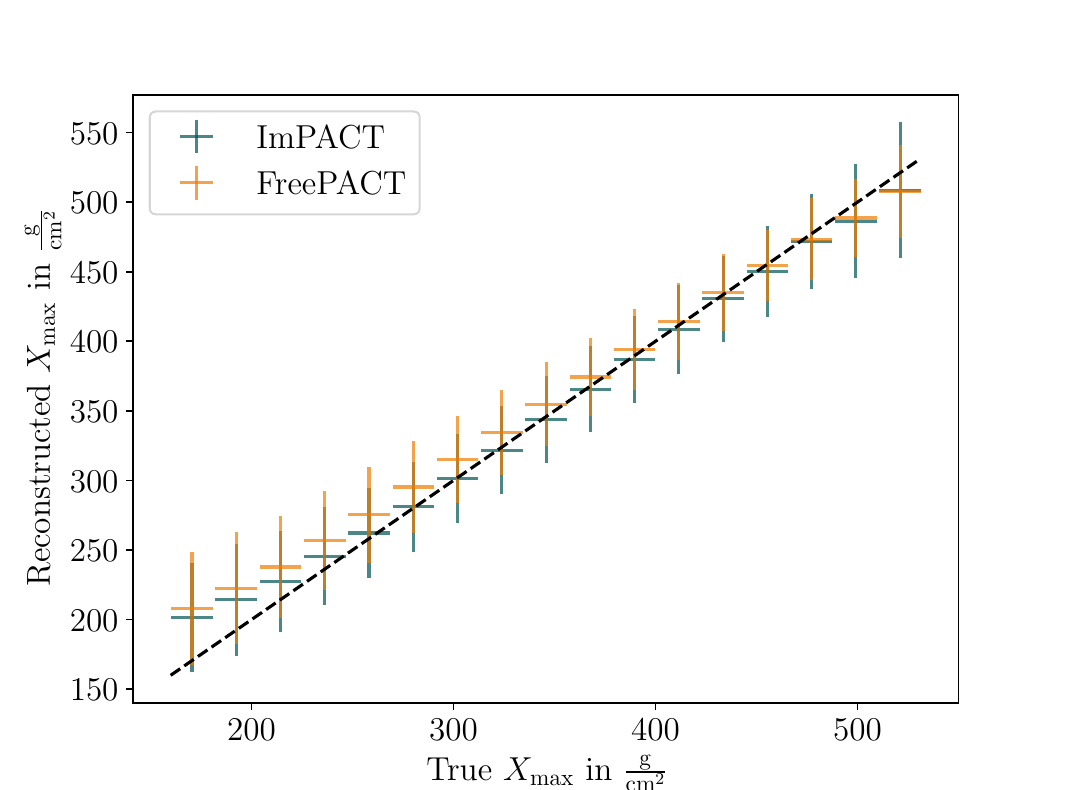}

\caption{
Reconstruction of $X_{\rm max}$ as achieved with the \emph{ImPACT} and \emph{Free\-PACT} algorithms on the full CTA southern {\it Alpha} configuration array.
}
\label{fig:xmax_reconstruction}
\end{figure}

In Figure~\ref{fig:xmax_reconstruction}, we show the resolution of the reconstruction of $X_{\rm max}$ for \emph{ImPACT} and \emph{Free\-PACT} for on-axis gamma rays at a zenith angle of $20^\circ$. There is no significant qualitative difference between the two methods. The low-$X_{\rm max}$ upward bias corresponds to the typical upward energy bias seen towards the low energy threshold of the telescope array. Given that a truly precise determination of $X_{\rm max}$ is of a generally secondary interest (e.g. for gamma-hadron separation) as it serves mostly as a nuisance parameter to account for differences in shower development, both \emph{ImPACT} and \emph{Free\-PACT} perform satisfactorily here.

\subsection{NSB Dependence}

\begin{figure}[!ht]
\centering
\includegraphics[trim={0cm 1.5cm 0cm 3.2cm},clip,scale=0.49]{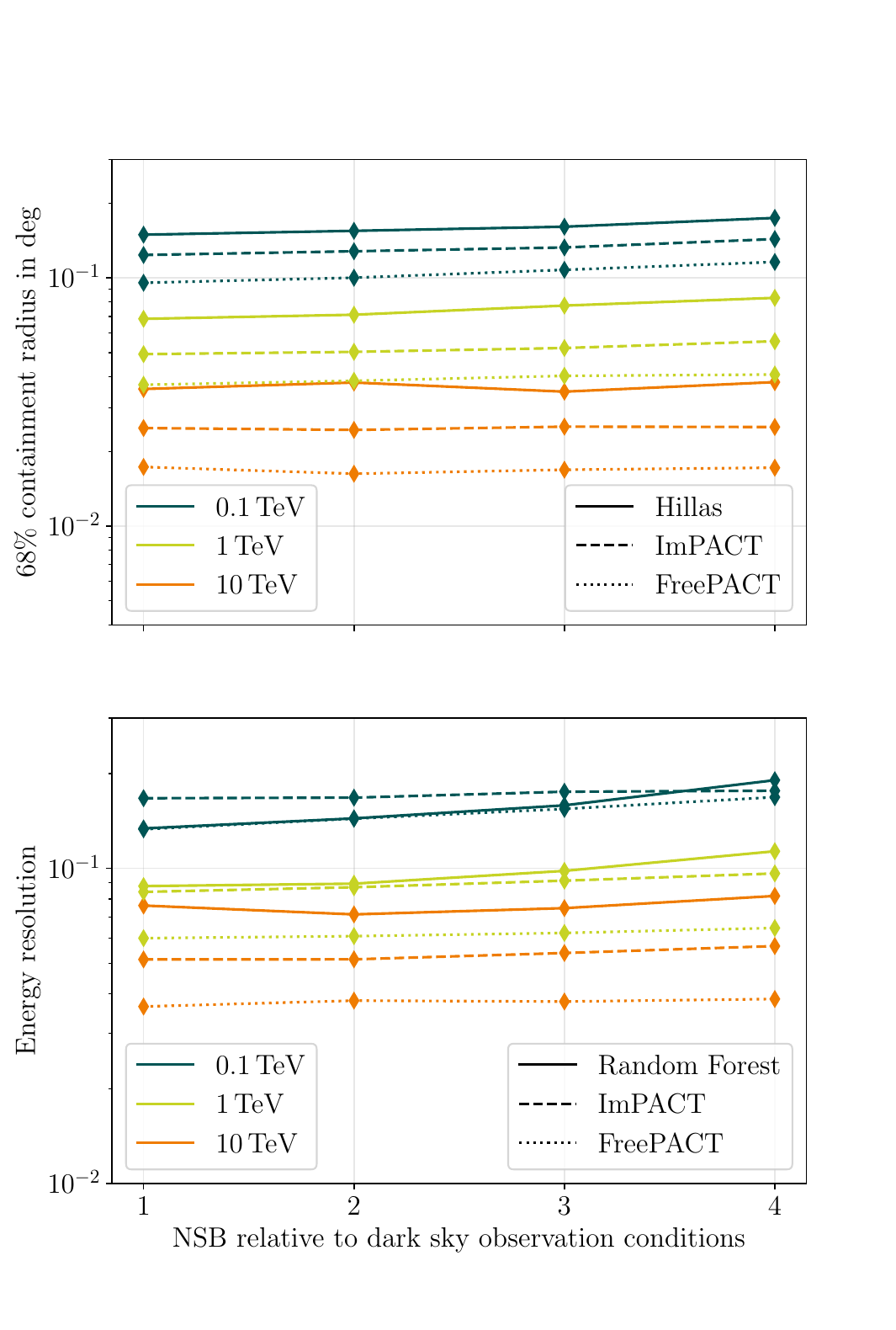}

\caption{
Angular and energy resolution for the full CTA southern array for on-axis photons at $20^{\circ}$ zenith angle as a function of NSB level. The solid lines show the performance of the standard Hillas/Random Forest analysis, the dashed lines show the results from a conventional \emph{ImPACT} analysis and the \emph{Free\-PACT} resolution curves are shown as dotted lines. Dark green, light green and orange curves represent the resolution at $0.1\,\mathrm{TeV}$, $1\,\mathrm{TeV}$ and $10\,\mathrm{TeV}$, respectively.
}
\label{fig:nsb_dependence}
\end{figure}

To demonstrate the robustness of the \emph{Free\-PACT} me\-thod with respect to changes in the observation conditions, we have produced test simulations of on-axis gamma rays at a zenith angle of $20^\circ$ for different levels of homogeneous NSB. We process and reconstruct these simulations with the exact same methods and settings as used for the simulations under default dark sky observation conditions. In particular, this means that we adjust neither the thresholds for the tail-cuts image cleanings nor the minimum image intensity requirement of $50$ P.E. Also, we do not adjust any of the reconstruction models to the different observation conditions, i.e. we do not train a new \emph{Free\-PACT} model or Random Forest energy regressor and do not adjust the pedestal width parameter in the analytical \emph{ImPACT} likelihood. Naturally, any of these changes would likely only improve the performance for high levels of NSB. In that sense, we consider here the worst-case scenario for the analysis of high NSB data. Keeping this in mind, we find in Figure \ref{fig:nsb_dependence} that the angular and energy resolutions degrade only very mildly with increased NSB for all algorithms and energies. In particular, the performance of the \emph{Free\-PACT} method remains superior to that of the Hillas/Random Forest and \emph{ImPACT} methods for all levels of NSB tested, and appears to be at least as robust as these methods with respect to changes in the observation conditions.

\subsection{Field-of-view Offset Dependence}
\begin{figure}[!t]
\centering
\includegraphics[trim={0cm 1.5cm 0cm 3.2cm},clip,scale=0.49]{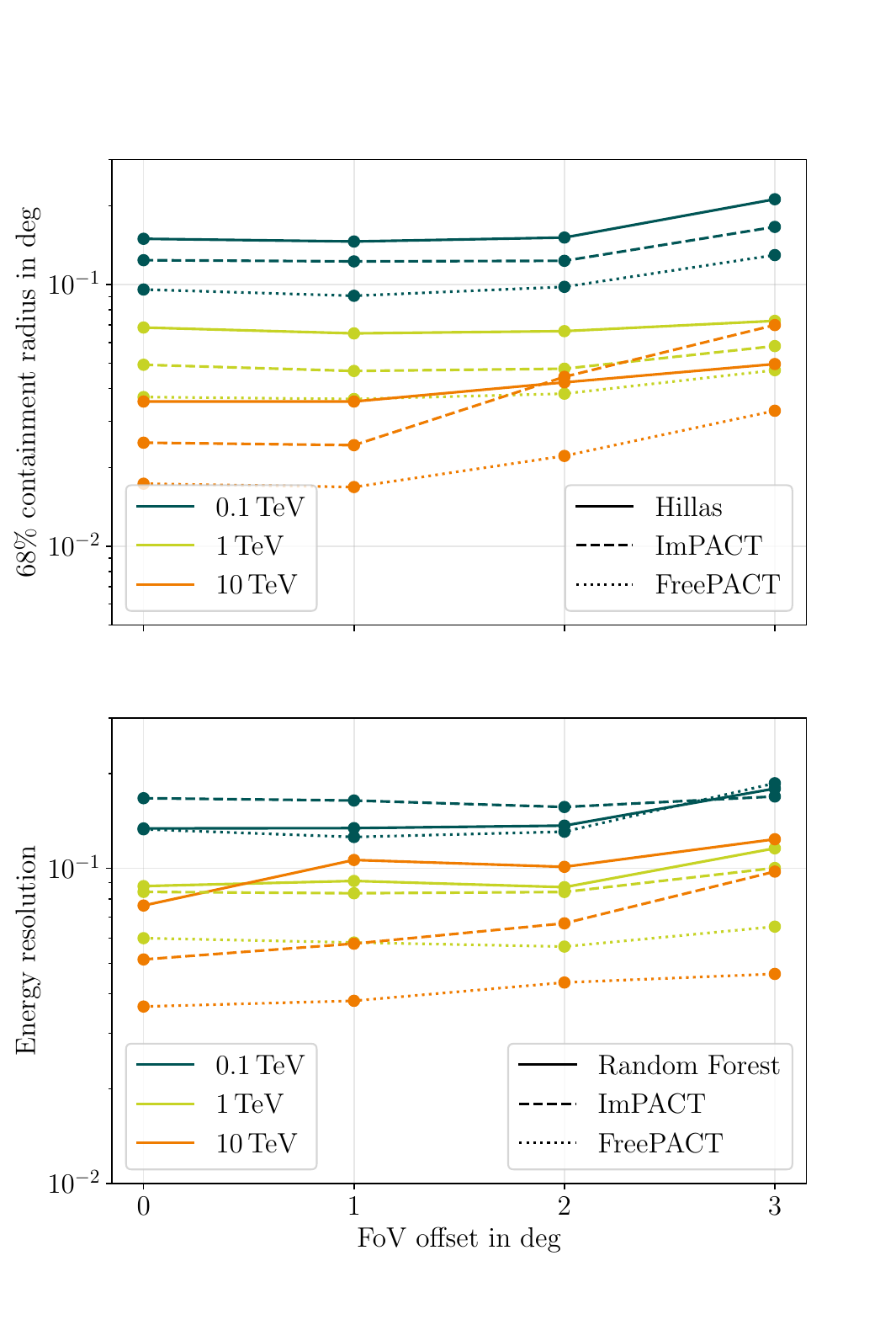}

\caption{
Angular and energy resolution for the full CTA southern array for $20^{\circ}$ zenith angle as a function of FoV offset. The solid lines show the performance of the standard Hillas/Random Forest analysis, the dashed lines show the results from a conventional \emph{ImPACT} analysis and the \emph{Free\-PACT} resolution curves are shown as dotted lines. The same \emph{ImPACT} templates, \emph{Free\-PACT} models and Random Forest regressor is used for the reconstruction at all FoV offsets. Dark green, light green and orange curves represent the resolution at $0.1\,\mathrm{TeV}$, $1\,\mathrm{TeV}$ and $10\,\mathrm{TeV}$, respectively.
}
\label{fig:offset_dependence}
\end{figure}

All results shown in this section thus far were derived from on-axis point source gamma rays. To assess the dependence of the reconstruction performance on field-of-view offset, we show in Figure~\ref{fig:offset_dependence} the angular and energy resolution for point source gamma rays for different source offsets from the array pointing directions. ("FoV offsets"). Similar to our test of NSB dependence, we make no changes to our event selection or reconstruction models from the on-axis case, with the exception of the Random Forest energy regressor which is now trained on diffuse gamma-ray simulations to provide a better seed for the likelihood fits. This once again comes close to a worst-case scenario for the analysis at large field-of-view offsets. Therefore, as expected, we find performance degradation of similar magnitude towards the edge of the field of view for all reconstruction methods. However, when comparing the reconstruction methods, we still find that the \emph{Free\-PACT} method performs best at all field-of-view offsets. It should be mentioned here that there are additional ways to better account for the field-of-view offset within the \emph{Free\-PACT} method that could be explored in future work. For example, the field-of-view offset of the gamma-ray source could become an additional input parameter to the \emph{Free\-PACT} models. Yet, even in the simplest case shown here, \emph{Free\-PACT} performs satisfactorily at large field-of-view offsets.

\subsection{Fit uncertainties}
\label{sec:fit_uncertainties}
\begin{figure*}[!ht]
\centering
\includegraphics[width=1.1\textwidth]{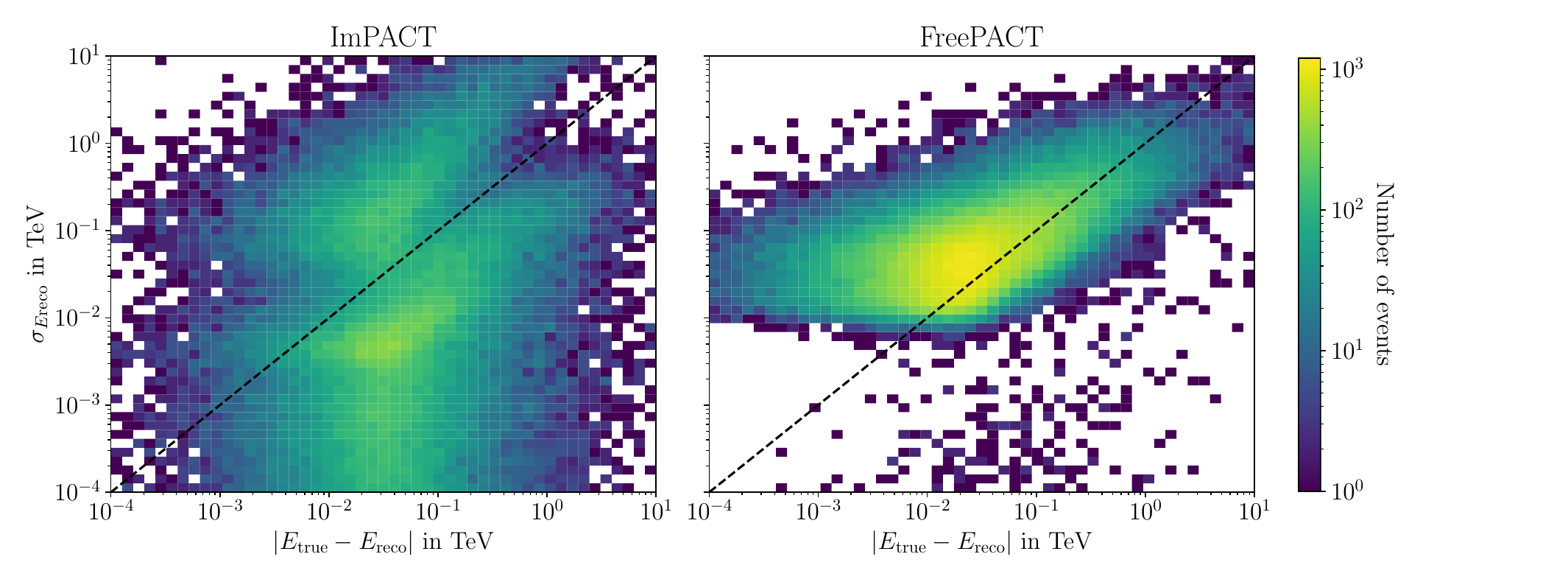}
\caption{
Correlation of difference between true and reconstructed gamma-ray energy with the reconstruction uncertainty from the \emph{Free\-PACT} and \emph{ImPACT} likelihood fits. The black dashed line indicates $\sigma_{E\mathrm{reco}}=|E_{\mathrm{true}}-E_{\mathrm{reco}}|$
}
\label{fig:uncertainty_correlation}
\end{figure*}

The uncertainties on the shower parameters determined in the likelihood fit provide another measure for the improved quality of the likelihood fit with the \emph{Free\-PACT} algorithm.
In Figure~\ref{fig:uncertainty_correlation}, we show the correlation between the difference between true and reconstructed energy and the uncertainty on the energy determined in the likelihood fit for \emph{ImPACT} and \emph{Free\-PACT}. These results are once again derived from on-axis events at a zenith angle of $20^\circ$ selected according to the criteria specified in Section~\ref{sec:event_selection}. Figure~\ref{fig:uncertainty_correlation} shows a significantly greater correlation for \emph{Free\-PACT} than for \emph{ImPACT}. This is a clear indication that \emph{Free\-PACT} not only provides a more accurate minimum of the likelihood than \emph{ImPACT}, but also a better description of the shape of the likelihood surrounding the minimum. Figure~\ref{fig:uncertainty_correlation} also illustrates that for \emph{Free\-PACT}, the fit uncertainty can be used to identify events for which the likelihood fit did not properly converge or got stuck in a local minimum. These events can be found in the tail of the distribution towards low fit uncertainties. They can be effectively removed from the event selection using a simple cut in $\sigma_{\mathrm{Ereco}}$. More sophisticated cuts, for example as a function of reconstructed energy, or a reprocessing of the events, possibly through a refit with a different seed, are of course also possible. These measures will in any case further improve the resolution achievable with \emph{Free\-PACT} relative to the curves shown in Figure~\ref{fig:full_cta_resolution curves}. 

\begin{figure*}[!ht]
\centering
\includegraphics[width=1.1\textwidth]{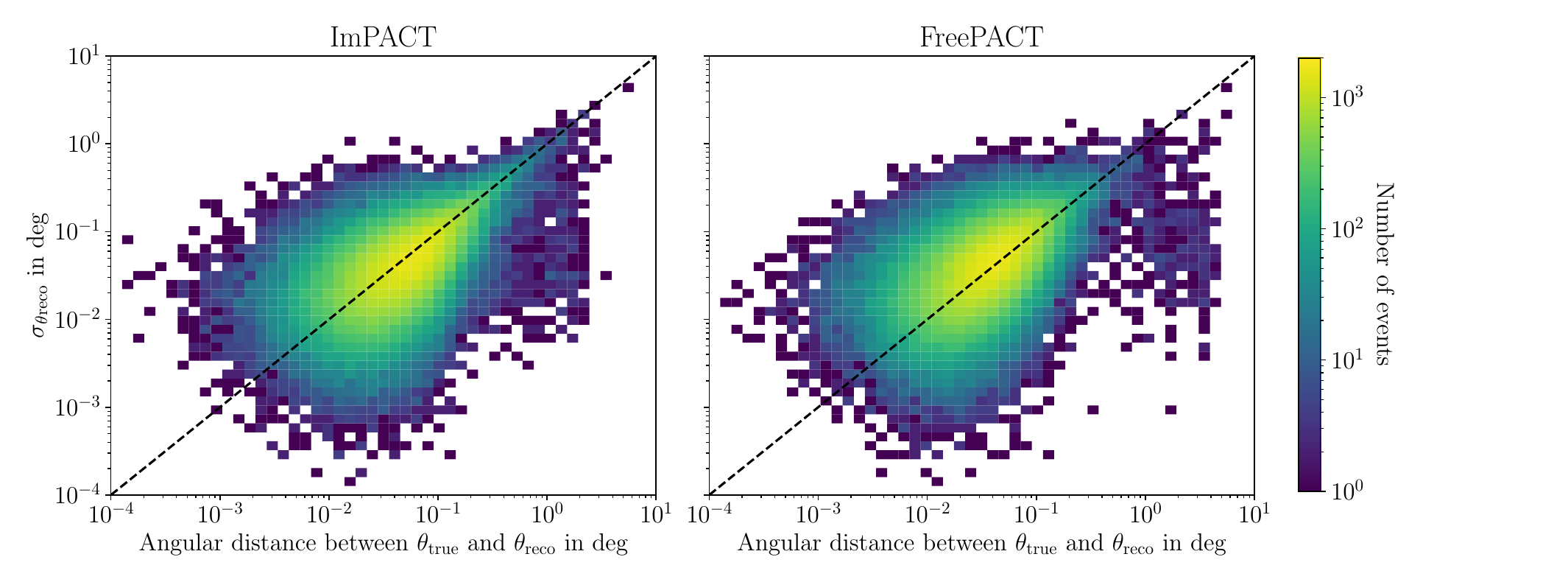}
\caption{
Correlation of difference between true and reconstructed gamma-ray source direction with the reconstruction uncertainty from the \emph{Free\-PACT} and \emph{ImPACT} likelihood fits. The black dashed line indicates where the reconstruction uncertainty $\sigma_{\theta\mathrm{reco}}$ is equal to the angular distance between true and reconstructed source direction.
}
\label{fig:angular_uncertainty_correlation}
\end{figure*}

For the same events as used for Figure~\ref{fig:uncertainty_correlation}, we compare in Figure~\ref{fig:angular_uncertainty_correlation} the correlation between the angular distance between true and reconstructed source direction and the uncertainty on the reconstructed source direction. We calculate the latter by adding in quadrature the uncertainties on the reconstructed altitude and azimuth coordinates. In this, the uncertainty on the reconstructed azimuth is corrected to take into account the zenith angle of the observation. Whereas there are major differences in the uncertainty on the reconstructed energy between \emph{ImPACT} and \emph{Free\-PACT}, we find that both \emph{ImPACT} and \emph{Free\-PACT} show a high level of correlation and thus provide meaningful uncertainties. The main differences between the distributions displayed in the left and right panels of Figure~\ref{fig:angular_uncertainty_correlation} lie in the size of the high angular distance tail which is smaller for \emph{Free\-PACT} than for \emph{ImPACT} and in the peak of the distribution that is shifted to both smaller angular distances and fit uncertainties for \emph{Free\-PACT}.

\subsection{Reconstruction speed}

Finally, besides the improvements in resolution presented above, the \emph{Free\-PACT} method also provides a significant speed up with respect to the \emph{ImPACT} method. Ultimately, this is because of two reasons: The evaluation of the \emph{Free\-PACT} model is faster than the interpolation on the grid of \emph{ImPACT} templates and the minimiser requires fewer iterations to find the minimum for \emph{Free\-PACT}. The precise speed-up factor typically lies between two and eight, and depends on the minimiser settings, whether the full convolved or the Gaussian \emph{ImPACT} likelihood is used and the size of the \emph{Free\-PACT} models.

\section{Discussion}
\label{sec:discuss}
The performance improvements provided by the \emph{Free\-PACT} method as described in Section~\ref{sec:perf} have implications for the scientific potential of IACT arrays in general and CTA specifically. Particularly interesting for galactic high-energy astrophysics are the performance improvements above gamma-ray energies of $\geq 50\,\mathrm{TeV}$ as the gamma rays in this energy range are produced in interactions of cosmic rays with energies of $\geq 500\,\mathrm{TeV}$. Currently, it is not fully understood which objects and environments in the Milky Way are responsible for the acceleration to these energies, though a number of candidate source classes have been hypothesised~\cite{Cristofari:2021jkl}. The number of detected gamma-ray sources with emission above $\geq 50\,\mathrm{TeV}$ has also increased significantly in recent years, in particular by water Cherenkov detectors such as HAWC~\citep{HAWC:2019tcx} and LHAASO~\citep{LHAASO:2023rpg}. These detectors can however not reach the resolution achievable with IACT arrays, limiting the interpretability of these discoveries. Therefore, sub-arcminute angular resolution and an energy resolution down to $3\%$ as achievable with the \emph{Free\-PACT} reconstruction for the CTA southern array allows for truly unprecedented spectro-morphological studies and a significantly more detailed understanding of these sources and the acceleration of galactic cosmic rays to PeV energies.

There are also still a few options to improve upon the \emph{Free\-PACT} method as presented here. In particular, the timing information of each pixel is currently not considered. In much the same way as presented here for the pixel charge, it would also be possible to train a neural network to learn the pixel hit time PDF $p(t_{\mathrm{hit}}|\vec{r}_{\rm pix},\vec{r}_{\rm tel},\vec{\eta})$. Then, charge and hit time likelihoods can be multiplied to form a combined likelihood. Another option that also accounts for the correlation between charge and hit time in a given pixel is to directly train a network to learn the combined charge and hit time likelihood $p(c,t_{\mathrm{hit}}|\vec{r}_{\rm pix},\vec{r}_{\rm tel},\vec{\eta})$. A further possibility is to not consider the hit times of individual pixels, but rather the time gradient of the pixels across the image and to learn a corresponding per-camera likelihood. 
The low-energy performance could also potentially be improved by relaxing the assumption of rotational symmetry of the telescopes and accounting for the orientation of the shower impact point from the telescope relative to the local geomagnetic field direction. This can likely be most easily done by including the angle between geomagnetic north and the line from telescope location to the shower impact point as a further parameter in the likelihood and the training of the \emph{Free\-PACT} models. This could be of particular relevance for the application of \emph{FreePACT} to the CTA northern array given the larger magnetic field at the site~\citep{Hassan:2017paq} and the presence of the Large-Sized Telescopes (LSTs) enhancing the sensitivity at the low-energy end.

Another aspect of the \emph{Free\-PACT} method that warrants a brief discussion is its potential for improving gam\Hyphdash ma-hadron separation. There are a few potentially useful parameters provided by the algorithm that could serve as input to decision tree based classifier methods. First, there are the fit uncertainties discussed in~\ref{sec:fit_uncertainties}. that are now more informative as for \emph{ImPACT}. Then, it is also possible to formulate a goodness-of-fit in much the same way as described in~\cite{Parsons:2014voa} for \emph{ImPACT}. Given the more accurate likelihood description achieved with the \emph{Free\-PACT} method, this parameter should become an even more powerful discriminant. 
These parameters can also be of further use elsewhere, for example for the generation of classes of events with different reconstruction quality as foreseen for CTA~\cite{CTAConsortium:2023ynz}. 

 Finally, all of the results presented in this work are derived from simulations of air showers and telescopes. Other CNN-based methods have in the past shown problems in translating the performance achieved on simulations to real data (see e.g.~\cite{Shilon:2018xlp}). For \emph{Free\-PACT}, we foresee this to be a lesser problem. This is because - very different from their role in the CNN-based methods- the role of the neural network in \emph{Free\-PACT} is just to approximate the charge PDF $p$. It is thus just as succeptible as any other likelihood-based reconstruction method to mismatches in this PDF between data and simulations. However, from the fact that \emph{ImPACT} translates well from simulations to data~\cite{Parsons:2014voa}, we know that the mean of the simulated PDF as contained in the \emph{ImPACT} templates has to agree well between data and simulations. Given that the width of the simulated PDF is set by either the calibrated pedestal charge distribution (noise-dominated regime) or well-understood electromagnetic physics responsible for the shower-to-shower fluctuations (signal-dominated regime), there is also no reason to expect significant deviations between the shape of this simulated PDF to that for real data. Additionally, there is also the intriguing possibility to improve the \emph{Free\-PACT} description of the charge likelihood under real observation conditions using the pixel noise level monitoring data that will be recorded by CTA. There are a couple of different ways in which this could be done: As one option, one could train a \emph{Free\-PACT} model on noiseless simulations and then convolve the resulting PDFs with the measured noise distributions. Another option is to include the RMS of simulated noise distributions as a further explicit (nuisance) parameter in the training of the \emph{Free\-PACT} models and then use the measured noise level in the evaluation and reconstruction. Naturally however, until it is actually applied to real IACT data, the translation of the performance of the \emph{Free\-PACT} reconstruction can not be fully judged. Therefore, the implementation of this method in the H.E.S.S. analysis framework and its subsequent application to H.E.S.S. data is foreseen.

\section{Summary and Conclusion}
\label{sec:summary}
We have presented the \emph{Free\-PACT} algorithm that improves upon previous image likelihood-fitting reconstruction techniques, in particular the \emph{ImPACT} method, for IACTs by replacing the analytical pixel charge PDF with the output of a neural network that approximates this PDF.

For the planned southern array of CTA, we have found this method to yield improvements of $50\%$ or more over wide energy range for both angular and energy resolution compared to \emph{ImPACT} and an analysis based on Hillas geometry reconstruction and a Random Forest energy regression method. We have demonstrated that this improvement is due to an improved description of the shape of the charge PDF in every camera pixel. This is also reflected in a more accurate estimation of the reconstruction uncertainties from the fit.

Besides the resolution improvements, the \emph{Free\-PACT} method also possesses significant operational advantages compared to the \emph{ImPACT}: No dedicated simulations are required for the training of the \emph{Free\-PACT} models, and  the reconstruction is sped up significantly as the evaluation of the \emph{Free\-PACT} models takes less time than the interpolation of the \emph{ImPACT} templates.

The improved resolution achievable with the \emph{Free\-PACT} method will ultimately help realise the full scientific potential of the CTA observatory and will allow for unprecedented precision in the study of the sources in the very high-energy gamma-ray sky.

\section*{Acknowledgements}

 This work was conducted in the context of the CTA Consortium. We thank the \emph{ctapipe} team, especially Maximilian Linhoff, Karl Kosack and Tomas Bylund, for their advise and assistance with the implementation of \emph{ImPACT} in \emph{ctapipe}. G.S. acknowledges membership in the International Max Planck Research School for Astronomy and Cosmic Physics at the University of Heidelberg (IMPRS-HD).
\bibliographystyle{elsarticle-num} 
\bibliography{library.bib}
\begin{appendix}
\section{CTA southern array layout in the {\it Alpha} configuration}
\label{app:array_layout}
\begin{figure}[H]
\centering
\includegraphics[trim={0cm 1cm 0cm 1cm},clip,scale=0.49]{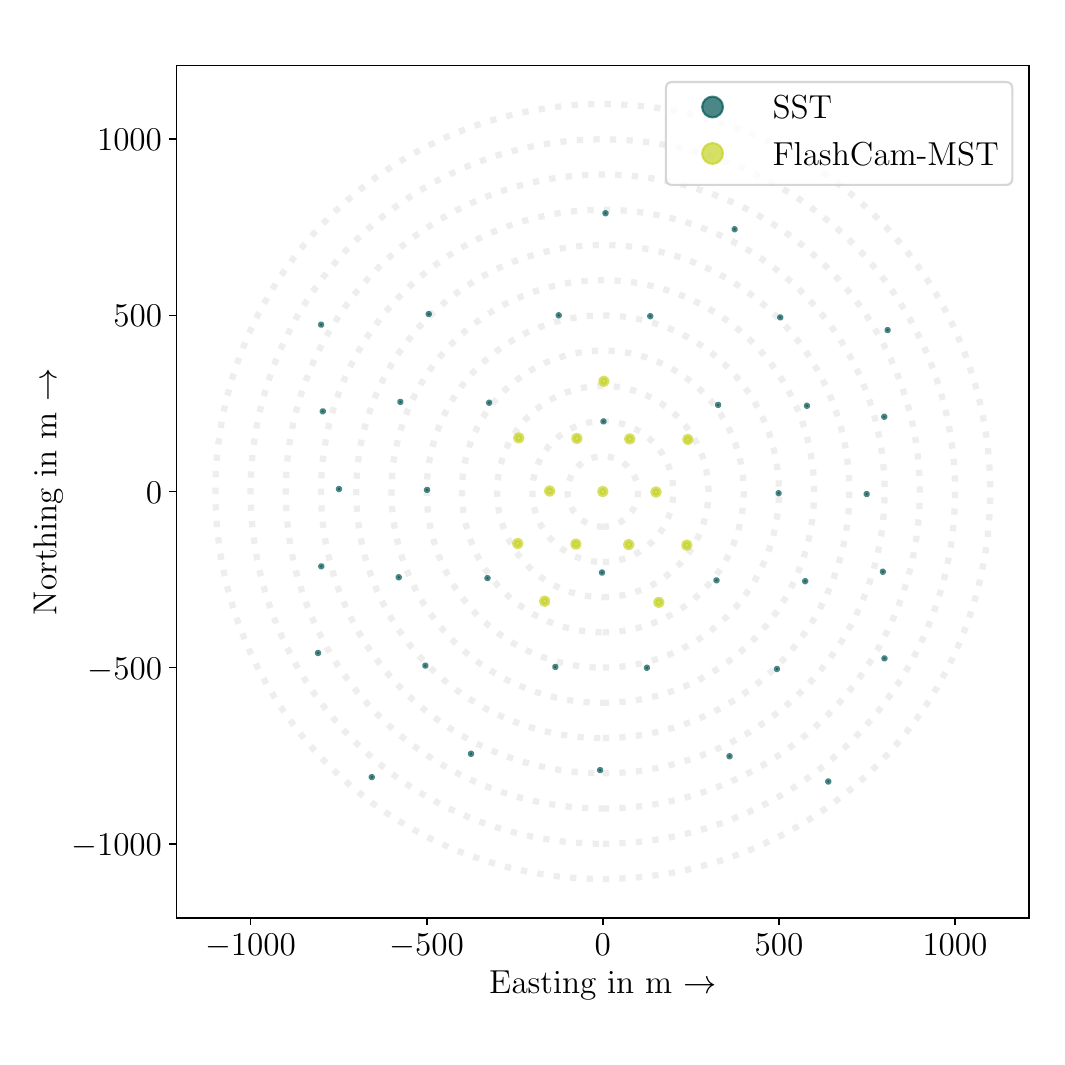}

\caption{
Telescope positions for the proposed {\it Alpha} configuration~\cite{alpha_config} of the CTA southern array consisting of 14 MSTs and 37 SSTs. To guide the eye, the dotted lines visualise circles of constant distance from the center of the array. The distance between neighboring lines is $100\,\mathrm{m}$.
}
\label{fig:array_layout}
\end{figure}

\section{SST Charge PDF Comparisons}
\label{app:sst_charge_pdf}

\begin{figure*}[!htbp]
\centering
\includegraphics[trim={0cm 0cm 0cm 1.5cm},clip,width=0.49\textwidth]{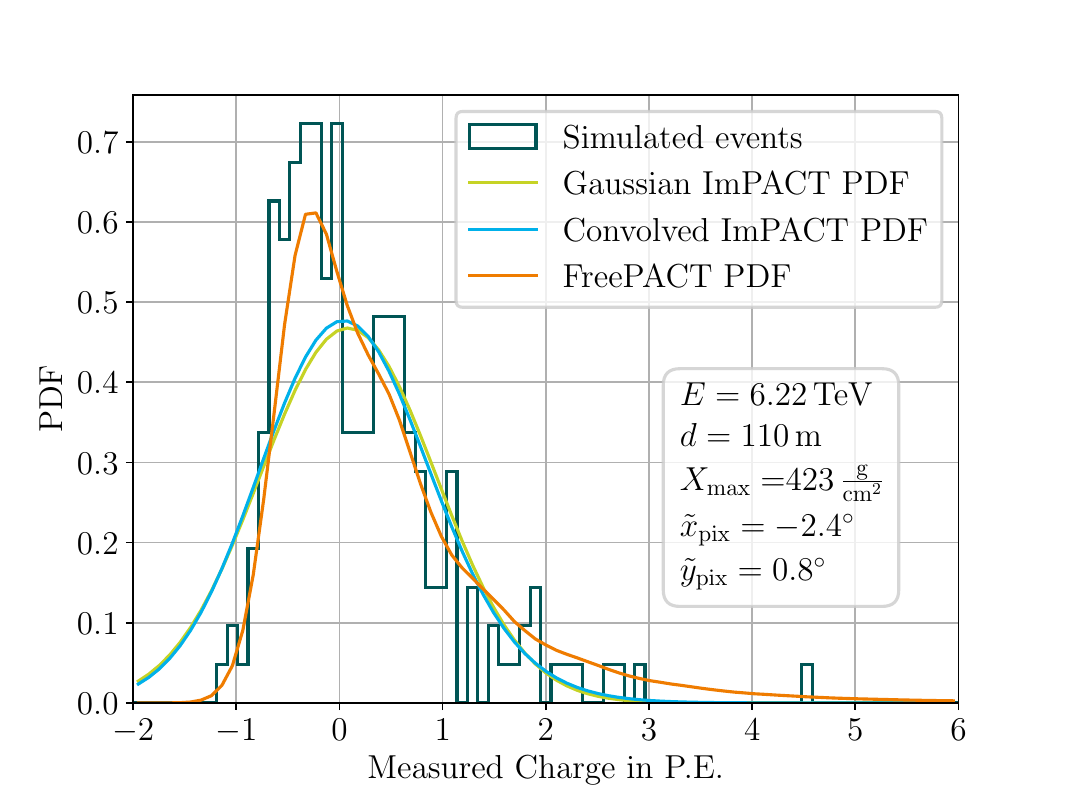}
\includegraphics[trim={0cm 0cm 0cm 1.5cm},clip,width=0.49\textwidth]{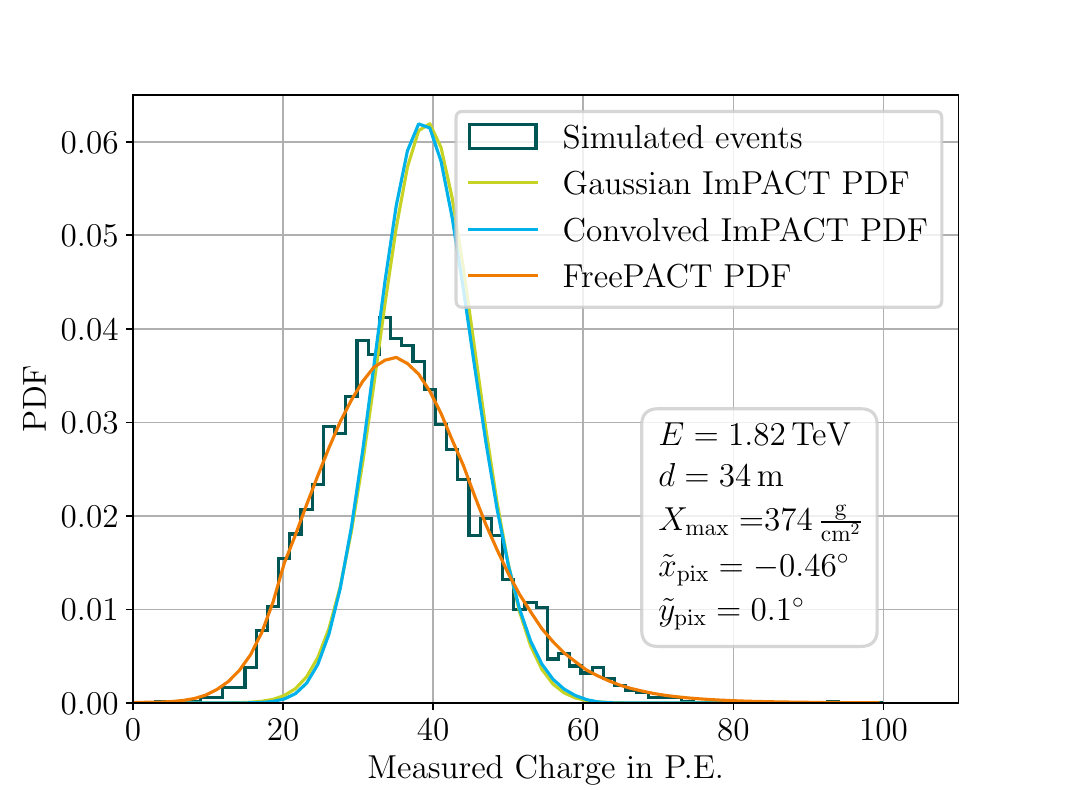}
\caption{
Charge distribution at the given pixel position for the given shower parameters in an SST with the analytical \emph{ImPACT} charge PDFs and the \emph{Free\-PACT} distribution. On the left, the charge is dominated by noise photons, on the right, the charge is dominated by Cherenkov photons
}
\label{fig:sst_charge_distribution}
\end{figure*}
In analogy to the MST results shown in Section~\ref{sec:charge_pdf_comparison}, we show here comparisons of the \emph{ImPACT} and \emph{Free\-PACT} charge PDFs to the distributions obtained from simulated events for the SST. The conclusions drawn from this are the same as for the MST: For pixels dominated by (electronics and NSB) noise, the \emph{Free\-PACT} PDF captures the specific shape of the noise distribution better than the analytical \emph{ImPACT} PDFs. For high-intensity pixels, the \emph{ImPACT} PDFs are too narrow as they do not account for shower-to-shower fluctuations, whereas the \emph{Free\-PACT} distribution gives a good description of the simulated distribution.

\section{Resolution curves after gamma-hadron separation}
\label{app:rel_curves_gh}
\begin{figure}[!ht]
\centering
\includegraphics[trim={0cm 3cm 0cm 4.8cm},clip,scale=0.49]{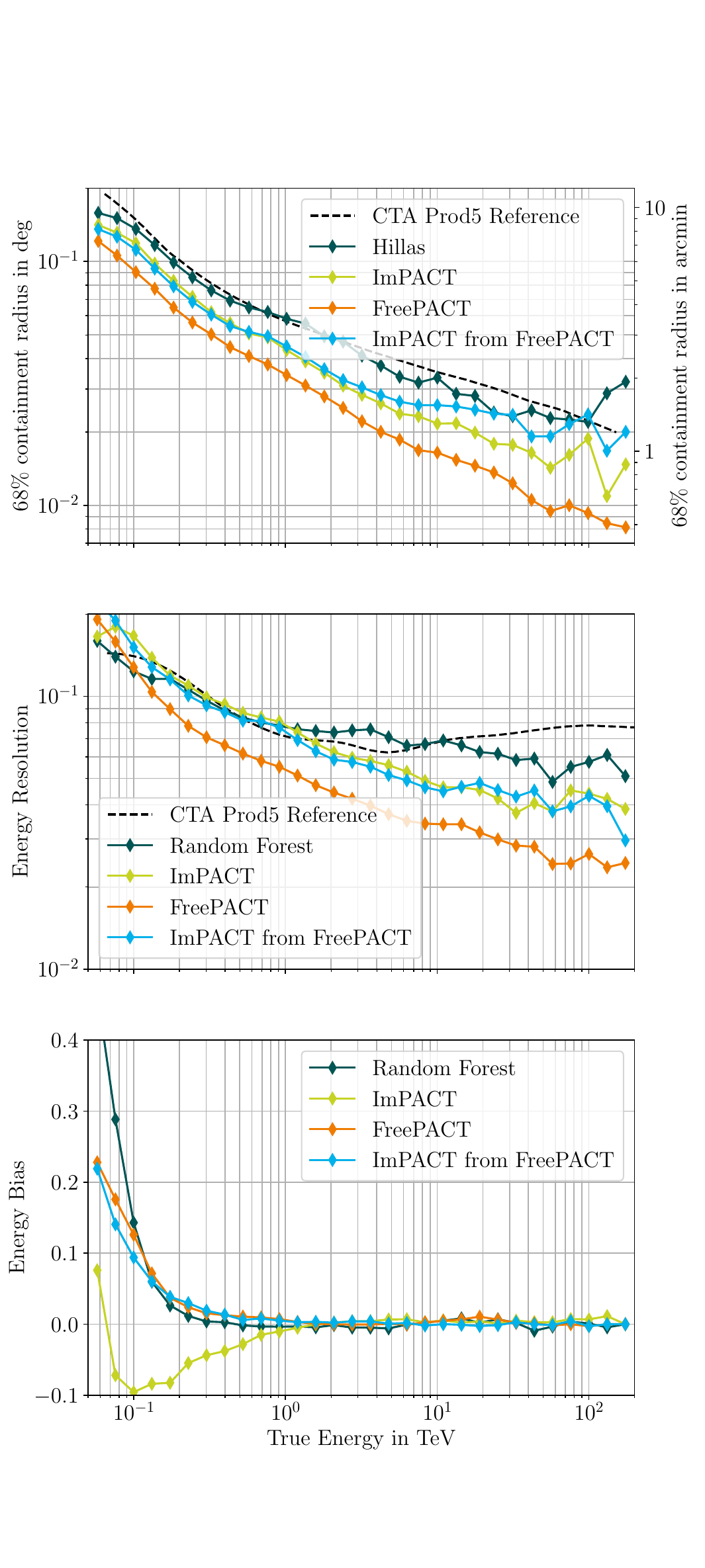}

\caption{
Angular resolution, energy resolution, and energy bias as a function of true gamma-ray energy for the full CTA southern array for on-axis photons at $20^{\circ}$ zenith angle after the application of a $75\%$ gamma-ray efficiency gamma-hadron separation cut. Shown in dark green is the standard Hillas/Random Forest analysis that serves as a seed to the image likelihood fits. In lime green, we show the results from a conventional \emph{ImPACT} analysis. The \emph{Free\-PACT} resolution curves are shown in orange. The light blue curve shows the resolution achieved with an \emph{ImPACT} analysis with templates generated from the \emph{Free\-PACT} models as described in Section \ref{sec:impact_templates}. The official CTA Prod5 curve (after gamma-hadron separation) is shown in black for reference.
}
\label{fig:full_cta_resolution curves_gh_cut}
\end{figure}

To estimate the effect of gamma-hadron separation on the resolution curves shown in Section \ref{sec:ang_E_res}, we train a Random Forest classifier on 14 image and reconstruction parameters. The reconstruction parameters are derived only from the Hillas/Random Forest analysis, and thus the exact same classifier can be used for all reconstruction methods. As we are only interested in a broad estimate of the effect of gamma-hadron separation, we choose the cut on the Random Forest classifier output such that $75\%$ of gamma rays are retained in 10 bins in reconstructed energy. The resulting resolution curves are shown in Figure~\ref{fig:full_cta_resolution curves_gh_cut}. Compared to Figure~\ref{fig:full_cta_resolution curves}, we find that both angular and energy resolution are slightly improved for all reconstruction methods. This is expected given that the gamma-hadron separation is supposed to filter out events that appear different from typical gamma ray events and are therefore also more likely to be misreconstructed. Overall, Figure \ref{fig:full_cta_resolution curves_gh_cut} demonstrates that our conclusions drawn in the main text are not affected by the application of gamma-hadron separation.

\section{Resolution curves for MST and SST sub-arrays}
\label{app:subarray_performance}

\begin{figure}[!ht]
\centering
\includegraphics[trim={0cm 3cm 0cm 4.8cm},clip,scale=0.49]{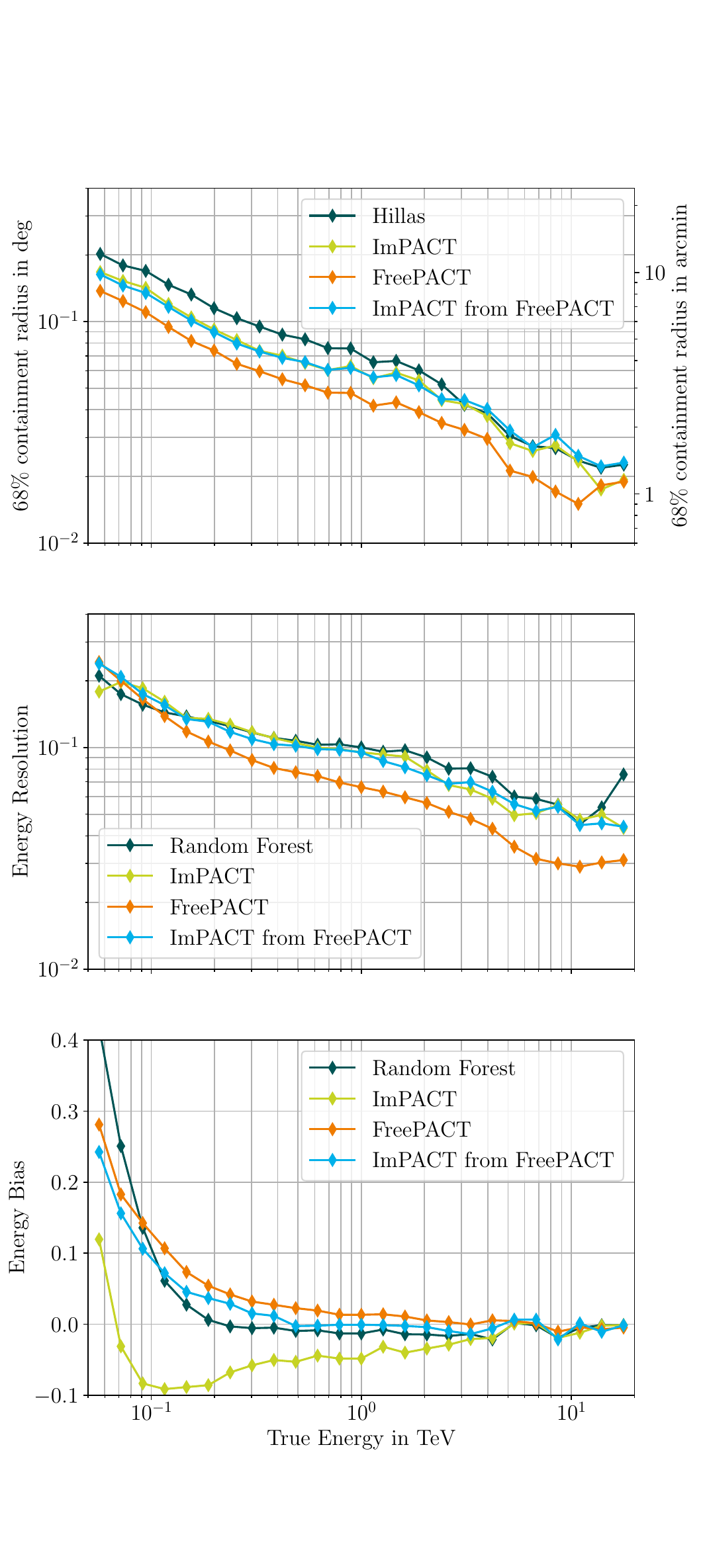}

\caption{
Angular resolution, energy resolution, and energy bias as a function of true gamma-ray energy for the 14 FlashCam-MST CTA southern array for on-axis photons at $20^{\circ}$ zenith angle. Shown in dark green is the standard Hillas/Random Forest analysis that serves as a seed to the image likelihood fits. In lime green, we show the results from a conventional \emph{ImPACT} analysis. The \emph{Free\-PACT} resolution curves are shown in orange. The light blue curve shows the resolution achieved with an \emph{ImPACT} analysis with templates generated from the \emph{Free\-PACT} models as described in Section \ref{sec:impact_templates}.
}
\label{fig:mst_resolution curves}
\end{figure}

\begin{figure}[!ht]
\centering
\includegraphics[trim={0cm 3cm 0cm 4.8cm},clip,scale=0.49]{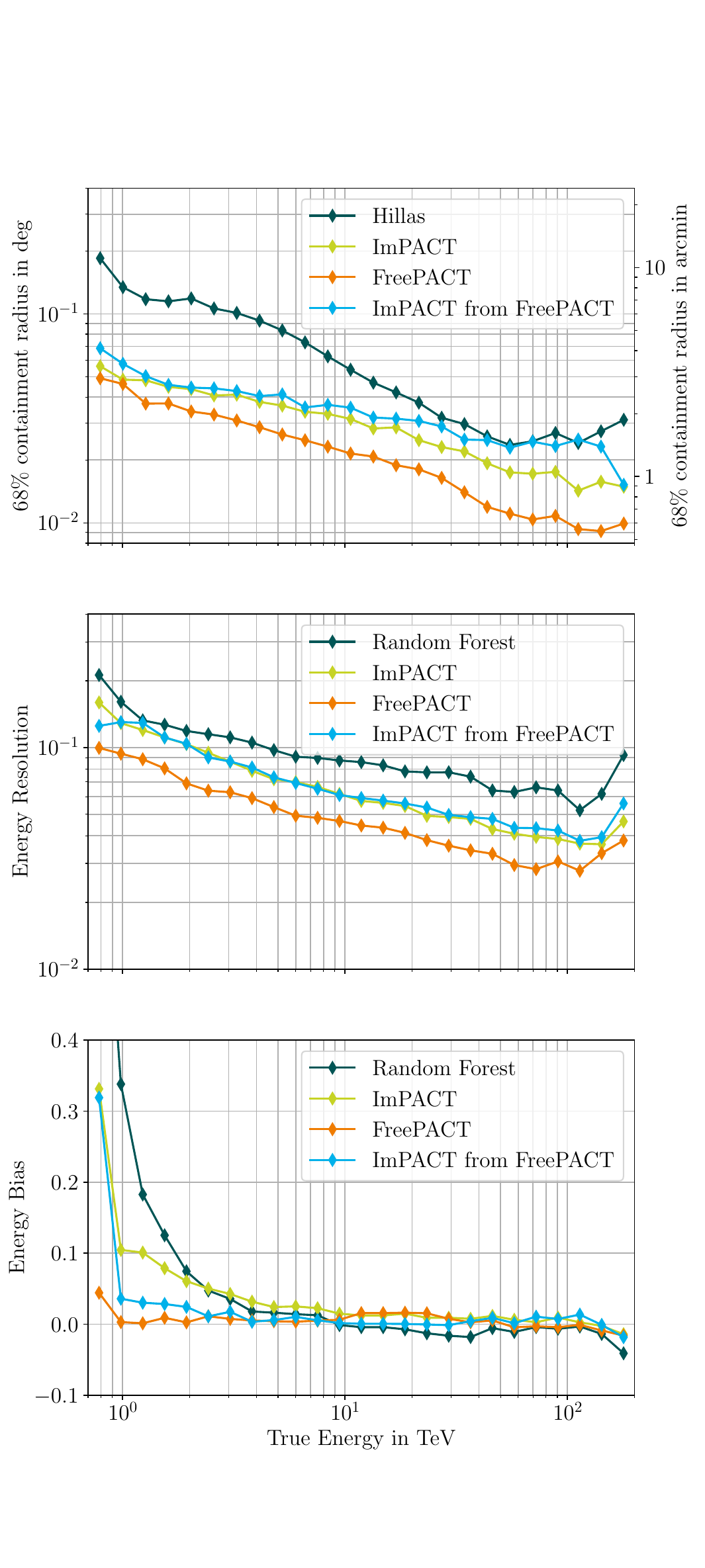}

\caption{
Angular resolution, energy resolution, and energy bias as a function of true gamma-ray energy for the 37 SST CTA southern array for on-axis photons at $20^{\circ}$ zenith angle. Shown in dark green is the standard Hillas/Random Forest analysis that serves as a seed to the image likelihood fits. In lime green, we show the results from a conventional \emph{ImPACT} analysis. The \emph{Free\-PACT} resolution curves are shown in orange. The light blue curve shows the resolution achieved with an \emph{ImPACT} analysis with templates generated from the \emph{Free\-PACT} models as described in Section \ref{sec:impact_templates}.
}
\label{fig:sst_resolution curves}
\end{figure}

In Figure~\ref{fig:mst_resolution curves} and Figure~\ref{fig:sst_resolution curves}, we show the energy and angular resolution curves for the MST and SST sub-array of the CTA southern array, respectively. The event selection in both cases is the same as described in section~\ref{sec:event_selection}, except that only three telescopes are required to pass the quality cuts instead of the four telescopes required for the full array.

As in Figure~\ref{fig:full_cta_resolution curves}, the results displayed are for on-axis point-source gamma rays at $20^{\circ}$ zenith angle. For the MST array, \emph{Free\-PACT} improves the angular resolution by more than $20\%$ over the entire energy range and the energy resolution by more than $30\%$ above $\approx 300\,\mathrm{GeV}$ relative to \emph{ImPACT}. For the SST array, the improvements are of the order of $50\%$ for angular resolution and $25\%$ for energy resolution over the entire energy range. These results are consistent with those for the full CTA southern array as presented in Section~\ref{sec:ang_E_res}. This is a good consistency check as the \emph{ImPACT} templates and \emph{Free\-PACT} model used for the analysis of the sub-arrays are exactly the same as used for the full array.

We also find very similar results for the two different sets of \emph{ImPACT} templates used for both telescopes, once again consistent with the findings for the full array.

\section{Resolution curves at \texorpdfstring{$50^\circ\,$}{z}zenith}
\label{app:fifty_zenith}
To demonstrate the performance of the \emph{Free\-PACT} me\-thod at zenith angles other than $20^\circ$, we show here as an example the resolution curves for on-axis observations at a zenith angle of $50^\circ$. Naturally, as the measured air shower images will differ from those at $20^\circ$ zenith angle, this necessitates the creation of a different set of \emph{ImPACT} templates as well as the training of different \emph{Free\-PACT} models and of a different Random Forest regressor. Other than that, the event selection is exactly the same as described for $20^\circ$ zenith angle in Section~\ref{sec:event_selection}. We find the angular and energy resolution to be improved by about $10\%$ with respect to \emph{ImPACT} at $300\,\mathrm{GeV}$. For the same comparison at $30\,\mathrm{TeV}$, we find the angular resolution to be improved by $20\%$ while the energy resolution is improved by $25\%$.

\begin{figure}[!ht]
\centering
\includegraphics[trim={0cm 3cm 0cm 4.8cm},clip,scale=0.49]{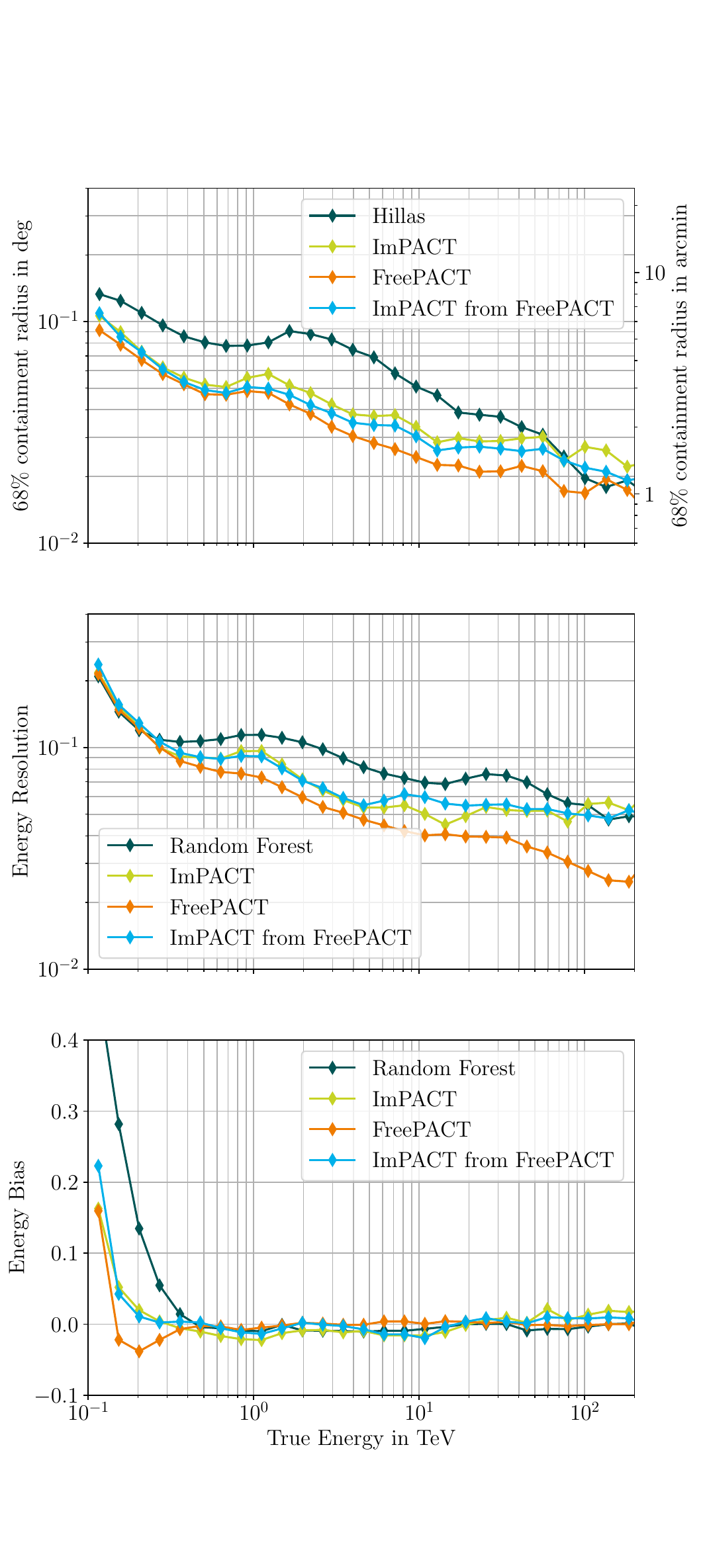}

\caption{
Angular resolution, energy resolution, and energy bias as a function of true gamma-ray energy for the full CTA southern array for on-axis photons at $50^{\circ}$ zenith angle. Shown in dark green is the standard Hillas/Random Forest analysis that serves as a seed to the image likelihood fits. In lime green, we show the results from a conventional \emph{ImPACT} analysis. The \emph{Free\-PACT} resolution curves are shown in orange. The light blue curve shows the resolution achieved with an ImPACT analysis with templates generated from the \emph{Free\-PACT} models as described in Section \ref{sec:impact_templates}.
}
\label{fig:full_cta_resolution curves_z50}
\end{figure}
\end{appendix}
\end{document}